\documentclass[11pt]{article}
\pdfoutput=1
\usepackage{jheppub}
\usepackage{caption}
\usepackage{subcaption}
\usepackage{amsfonts}
\usepackage{amssymb}
\usepackage{amsmath}
\usepackage{multirow}
\usepackage{color}
\usepackage{ulem}
\newcommand{\beq}{\begin{eqnarray}}
\newcommand{\eeq}{\end{eqnarray}}

\title{Search for a lighter Higgs Boson in Two Higgs Doublet Models}
\author[a]{Giacomo Cacciapaglia}
\author[a,1]{, Aldo Deandrea\note{also Institut Universitaire de France, 103 boulevard Saint-Michel, 75005 Paris, France}}
\author[a]{, Suzanne Gascon-Shotkin}
\author[a,2]{, Sol\`ene Le Corre\note{leading author}} 
\author[a]{, Morgan Lethuillier} 
\author[b]{, Junquan Tao} 
\affiliation[a]{Univ. Lyon, Universit{\' e} Claude Bernard Lyon 1, CNRS/IN2P3, UMR5822 IPNL, F-69622, Villeurbanne, France}
\affiliation[b]{Inst. High Energy Physics, Chinese Academy of Sciences, P.O. Box 918, Beijing 100049, China}
\emailAdd{g.cacciapaglia@ipnl.in2p3.fr}
\emailAdd{deandrea@ipnl.in2p3.fr}
\emailAdd{smgascon@in2p3.fr}
\emailAdd{s.le-corre@ipnl.in2p3.fr}
\emailAdd{morgan.lethuillier@cern.ch}
\emailAdd{taojq@ihep.ac.cn}
\abstract{We consider present constraints on Two Higgs Doublet Models, both from the LHC at Run 1 and from other sources in order to explore the possibility 
of constraining a neutral scalar or pseudo-scalar particle lighter than the 125 GeV Higgs boson. Such a lighter particle is not yet 
completely excluded by present data. We show with a simplified analysis that some new constraints could be obtained at the LHC if such a search 
is performed by the experimental collaborations, which we therefore encourage to continue carrying out light diphoton resonance searches at $\sqrt{s}=$ 
13 TeV in the context of Two Higgs Doublet Models.}

\begin{document}
 
 \maketitle
 
 \section{Introduction}

After the discovery of a Higgs boson at the LHC in 2012 \cite{Aad:2012tfa,Chatrchyan:2012xdj}, many studies, both from the theoretical and experimental 
side, have considered extensions of the Standard Model (SM) with an enlarged scalar sector. Concerning this scalar sector of physics beyond the Standard 
Model (BSM), most studies have considered the possibility of new scalars heavier than the 125~GeV Higgs boson which was discovered at the LHC. It is however
possible to have a spectrum in which lighter scalars are present together with an SM-like Higgs boson at 125~GeV. Among these possibilities there are  
detailed BSM models as well as effective descriptions including only the extended scalar sector. Two Higgs Doublet Models (2HDMs) constitute one of the simplest possibilities, where the SM Lagrangian is extended by the addition of a second scalar doublet. Previous phenomenological studies describing the possibility of lighter Higgs bosons include 
\cite{Ferreira:2012my,Chang:2012ve,Chang:2013ona,Celis:2013rcs,Cacciapaglia:2013ora,Bernon:2015wef,Bernon:2014nxa}, while for a recent study in supersymmetry (which naturally includes two doublets) we refer the reader to~\cite{Ellwanger:2015uaz}. At masses below 125 GeV, the main search channel at the LHC is the di-photon decay channel~\cite{CMS_diphoton,Aad:2014ioa}. 
This paper is organised as follows: in Section \ref{sec:2hdm} we describe the theoretical set-up adopted in our analysis; Section \ref{sec:bounds} is dedicated 
to the study of present constraints coming from flavour physics, electroweak precision tests, theoretical bounds, direct LEP constraints on the scalar sector and LHC limits given by a 125 GeV Higgs boson; Section \ref{sec:search} contains the 
cross section and branch ratio calculations for a light scalar Higgs boson, a study of the parameter space of the different 2HDMs and a comparison with the CMS low mass di-photon analysis at 8 TeV \cite{CMS_diphoton};
Section \ref{sec:pseudo} is dedicated to the study of the case where the lighter resonance is pseudo-scalar; finally we present our conclusions in Section \ref{sec:conclusions} .

\section{Two Higgs Doublet Models}
\label{sec:2hdm}
We here briefly describe the theoretical framework of 2HDMs, see \cite{2HDM_lecture} for a general discussion.
The 2HDMs are a simple extension of the Standard Model including two complex SU(2) doublets, $\phi_1$ and $\phi_2$. In order to avoid flavour-changing neutral 
currents, one can introduce a $\mathbb{Z}_2$ symmetry so that all fermions of a given electric charge couple to at most 
one Higgs doublet. These couplings can occur in different ways; the convention usually adopted is given in Table \ref{table:types}.

\begin{table}[h!]
  \begin{center}
    \begin{tabular}{c|c|c|c|c}
      & Type I & Type II & Flipped & Lepton Specific \\ 
      &            &             & (Type Y) & (Type X) \\ \hline
    Up-type quark & $\phi_2$ & $\phi_2$ & $\phi_2$ & $\phi_2$ \\ \hline
    Down-type quark & $\phi_2$ & $\phi_1$ & $\phi_1$ & $\phi_2$ \\ \hline
    Leptons & $\phi_2$ & $\phi_1$ & $\phi_2$ & $\phi_1$ \\ \hline
    \end{tabular}
  \caption{The different possible couplings between the SM fermions and the two scalar doublets in 2HDMs.}
  \label{table:types}
  \end{center}
\end{table}

The most generic 2HDMs potential constrained by the $\mathbb{Z}_2$ symmetry can be written as:
 \begin{align}
  V = & m_{11}^2 \phi_1^{\dagger}\phi_1 +m_{22}^2 \phi_2^{\dagger}\phi_2 - m_{12}^2 \left( \phi_1^{\dagger}\phi_2 + \phi_2^{\dagger}\phi_1 \right) + \frac{\lambda_1}{2} \left(\phi_1^{\dagger}\phi_1 \right)^2 + \frac{\lambda_2}{2} \left(\phi_2^{\dagger}\phi_2 \right)^2 + \nonumber \\ 
  & \lambda_3 \left( \phi_1^{\dagger}\phi_1 \right) \left( \phi_2^{\dagger}\phi_2 \right) + \lambda_4 \left( \phi_1^{\dagger}\phi_2 \right) \left( \phi_2^{\dagger}\phi_1\right) + \frac{\lambda_5}{2} \left[ \left( \phi_1^{\dagger}\phi_2 \right)^2 + \left( \phi_2^{\dagger}\phi_1 \right)^2 \right]\,,
  \label{eq:potential}
  \end{align}
 where all the parameters are real. The parameter $m_{12}^2$ is responsible for a soft breaking of the $\mathbb{Z}_2$ symmetry.
The two scalar doublets acquire vacuum expectation values (vevs):
 \begin{equation}
  \phi_1= \left( \begin{array}{c}
                  0 \\ \frac{v_1}{\sqrt{2}}
                 \end{array}
	  \right), \qquad
  \phi_2= \left( \begin{array}{c}
                  0 \\ \frac{v_2}{\sqrt{2}}
                 \end{array}
	  \right),
 \end{equation}
with $v \equiv \sqrt{v_1^2 + v_2^2}$.

After symmetry breaking we are left with five physical scalars: two neutral $\mathcal{CP}$-even states $h$ and $H$, one neutral $\mathcal{CP}$-odd state $A$ and two charged ones $H^{\pm}$. In order to move from the potential of eq.~\ref{eq:potential} to mass-eigenstates, one needs to introduce two angles: $\beta$, defined as $\tan \beta = \frac{v_2}{v_1}$, which rotates the two doublets in a basis where only one of them acquires a vev, and $\alpha$ which mixes the $\mathcal{CP}$-even scalar states to give mass-eigenstates. The parameters of the potential can thus be translated into an equivalent set of parameters in the physical basis:
 \begin{gather}
  \lambda_1,~ \lambda_2,~ \lambda_3,~ \lambda_4,~ \lambda_5, ~m_{11}^2, ~m_{22}^2, ~m_{12}^2 \nonumber \\
  \Updownarrow \\
  m_h, ~m_H, ~m_A, ~m_{H^{\pm}}, ~\tan\beta, ~\sin(\beta - \alpha),~ v, ~ m_{12}^2 \nonumber
 \end{gather}
where $v$ is set to the electroweak scale, and one of the masses of the $\mathcal{CP}$-even states should be equal to the measured Higgs boson mass. 
The masses of the two $\mathcal{CP}$-even states are ordered with $m_h < m_H$, where we will call $h$ the light Higgs boson and $H$ the heavy Higgs boson of the model. 
The couplings between the neutral Higgs bosons and the fermions and gauge bosons are summarised in Table \ref{table:couplings}.
In the rest of the study we will use the input parameters of the physical basis: we fix $v=246$ GeV and the heavy Higgs boson $H$ of the model 
is identified with the Higgs boson discovered at LHC, $m_H=125$ GeV, while the remaining six parameters are left free.  
 \begin{table}[h!]
  \begin{center}
    \begin{tabular}{c|c|c|c|c|c|}
      \cline{2-6}& & Type I & Type II & Flipped & Lepton Specific \\ \hline
       \multirow{3}{*}{Up-Type quark} & h & \multicolumn{4}{|c|}{$\frac{\cos\alpha}{\sin\beta}$} \\ 
       & H & \multicolumn{4}{|c|}{$\frac{\sin\alpha}{\sin\beta}$}\\
        & A & \multicolumn{4}{|c|}{$\cot\beta$} \\ \hline
        \multirow{3}{*}{Down-Type quark} & h & $\frac{\cos\alpha}{\sin\beta}$ & $-\frac{\sin\alpha}{\cos\beta}$ & $-\frac{\sin\alpha}{\cos\beta}$ & $\frac{\cos\alpha}{\sin\beta}$ \\
       & H & $\frac{\sin\alpha}{\sin\beta}$ & $\frac{\cos\alpha}{\cos\beta}$ & $\frac{\cos\alpha}{\cos\beta}$ & $\frac{\sin\alpha}{\sin\beta}$ \\
       & A & $\cot\beta$ & $\tan\beta$ & $\tan\beta$ & $\cot\beta$\\ \hline
      \multirow{3}{*}{Lepton} & h & $\frac{\cos\alpha}{\sin\beta}$ & $-\frac{\sin\alpha}{\cos\beta}$ & $\frac{\cos\alpha}{\sin\beta}$ & $-\frac{\sin\alpha}{\cos\beta}$ \\
      & H & $\frac{\sin\alpha}{\sin\beta}$ & $\frac{\cos\alpha}{\cos\beta}$ & $\frac{\sin\alpha}{\sin\beta}$ & $\frac{\cos\alpha}{\cos\beta}$ \\
      & A & $\cot\beta$ & $\tan\beta$ & $\cot\beta$ & $\tan\beta$ \\ \hline
      \multirow{3}{*}{WW and ZZ} & h & \multicolumn{4}{|c|}{$\sin(\beta-\alpha)$} \\
      & H & \multicolumn{4}{|c|}{$\cos(\beta-\alpha)$} \\
      & A & \multicolumn{4}{|c|}{0} \\ \hline
    \end{tabular}
  \caption{Tree level couplings between the neutral Higgs bosons and the gauge bosons and fermions normalised to their SM values for the different 2HDMs.}
  \label{table:couplings}
  \end{center}
 \end{table}


\section{Bounds on 2HDMs}
 \label{sec:bounds}
As we briefly discussed in the previous section, one of the simplest modifications of the SM consists in incorporating two scalar doublets, imposing custodial 
symmetry in order to allow satisfying the electroweak precision tests. The spectrum of neutral and charged scalars of the 2HDMs is a minimal extension of the 
scalar sector with one additional doublet and gives rise to five physical scalars: two charged $H^{\pm}$ and three neutral $h$, $H$ and $A$ states. If the Higgs 
boson discovered at the LHC is associated with the heavier $H$, the two other neutral states can be a candidate for a lighter Higgs boson.
The CMS collaboration has reported results on the search for a light resonance in di-photon final states \cite{CMS_diphoton}, giving the observed upper limit at 95\% confidence level (C.L.) on the cross section times branching ratio as a function of the mass of a light Higgs boson between 80 GeV and 110 GeV. 

In the following we list the different constraints we use to impose bounds on the model. We split them in three classes: indirect constraints, LEP constraints and LHC constraints.

\subsection{Indirect constraints}

The indirect constraints we apply on the 2HDMs parameter space include limits on the oblique parameters S, T and U \cite{Peskin:1991sw} due to electroweak precision tests, flavour constraints
and theoretical requirements due to ensure stability of the potential, unitarity and perturbativity.

The oblique parameters are computed in the model via the program \texttt{2HDMC} \cite{2HDMC} and compared to the experimental limits \cite{oblique_parameters} at 2$\sigma$ (see Table \ref{table:STU_param} for a recap of the updated experimental values with $1\sigma$ uncertainties and the correlations between them).
 \begin{table}[h!]
  \begin{center}
    \begin{tabular}{c|c}
        & Experimental values\\ \hline
       S & 0.05 $\pm$ 0.11 \\ \hline
       T & 0.09 $\pm$ 0.13 \\ \hline
       U & 0.01 $\pm$ 0.11 \\ \hline
    \end{tabular}
    \quad
    \begin{tabular}{c|c}
     & Correlations \\ \hline
     ST & +0.90 \\ \hline
     SU & -0.59 \\ \hline
     TU & -0.83 \\ \hline
    \end{tabular}
  \end{center}
  \caption{Experimental values of the oblique parameters with $1\sigma$ uncertainty and correlations between them \cite{oblique_parameters}.}
  \label{table:STU_param}
\end{table} 

The stability of the potential is needed in order to allow symmetry breaking with a stable vacuum, thus the potential of the theory needs to be bounded from below. 
 This condition requires \cite{2HDM_lecture}:
 \begin{gather}
  \lambda_1 \geq 0, \qquad \lambda_2 \geq 0, \qquad \lambda_3 \geq - \sqrt{\lambda_1 \lambda_2},\nonumber \\
  \lambda_3 + \lambda_4 - |\lambda_5| \geq - \sqrt{\lambda_1 \lambda_2}
 \end{gather}
 In addition we require to have tree-level perturbative unitarity for the scattering of Higgs bosons and the longitudinal parts of electroweak gauge bosons~\cite{Arhrib:2012ia}.\\
 In order to trust perturbative calculations, we add a condition on the quartic Higgs bosons couplings $C_{h_ih_jh_kh_l}$: 
 \begin{equation}
  |C_{h_ih_jh_kh_l}|\leq 4 \pi\
 \end{equation}
The three conditions detailed above are also computed via the \texttt{2HDMC} program.\\
 
 Once the previous requirements are satisfied, the available parameter space is tested against flavour bounds.
We look at the branching ratios  $\mathcal{BR}(\overline{B} \rightarrow X_s \gamma)$ and $\mathcal{BR}(B_s \rightarrow \mu^{+} \mu^{-})$, which obtain 
new contributions from the charged Higgs bosons and the neutral ones respectively and at the isospin asymmetry $\Delta_0(B \rightarrow K^*\gamma)$ 
and the $\Delta M_d$ frequency oscillation which are sensitive to the presence of charged Higgs bosons.
The value of each process is computed in the 2HDMs via the program \texttt{SuperIso} \cite{SuperIso,Mahmoudi:2007vz} and then compared to the experimental 
limits at $2\sigma$.
In order to take into account the theoretical uncertainties in the 2HDMs calculation, which are not evaluated in \texttt{SuperIso}, we add to the experimental $1 \sigma$ uncertainty $\sigma_{Exp}$ of each process the $1\sigma$ theoretical uncertainty $\sigma_{Th}$ of this same process computed in the SM given by the most 
recent theoretical calculations. The combined error $\sigma_{comb}$ can then be obtained via:
 \begin{equation*}
  \sigma_{comb}=\sqrt{\sigma_{Exp}^2+ \sigma_{Th}^2}
 \end{equation*}
A summary of the results we use is available in Table \ref{table:flavor_const}.

\begin{table}[h!]
  \begin{center}
    \begin{tabular}{c|c|c|c}
       Process & Experimental values & Theoretical computation & Combined error at $1\sigma$\\ \hline
       $\mathcal{BR}(\overline{B} \rightarrow X_s \gamma)$  & $(3.43 \pm 0.22)\times 10^{-4}$ \cite{Amhis:2014hma} & $(3.40 \pm 0.19)\times 10^{-4}$ \cite{Hurth:2016fbr} & $0.29\times10^{-4}$ \\ \hline
       $\mathcal{BR}(B_s \rightarrow \mu^{+} \mu^{-})$ & $(2.9 \pm0.7)\times10^{-9}$ \cite{Aaij:2013aka,CMS:2014xfa}& $(3.54\pm0.27)\times10^{-9}$ \cite{Hurth:2016fbr} & $0.8\times10^{-9}$\\ \hline
       $\Delta_0(B \rightarrow K^*\gamma)$ & $(5.2 \pm 2.6)\times 10^{-2}$ \cite{PDG} & $(5.1\pm1.5)\times10^{-2}$ \cite{Hurth:2016fbr} & $3.0\times10^{-2}$ \\ \hline
       $\Delta M_d$ & $0.510\pm0.003$ ps$^{-1}$ \cite{Amhis:2014hma} & $0.543\pm0.091$ ps$^{-1}$ \cite{Lenz:2014jya} & $0.091$ ps$^{-1}$\\ \hline
    \end{tabular}
  \end{center}
  \caption{Values of the experimental and theoretical flavour constraints.}
  \label{table:flavor_const}
\end{table}

\subsection{Direct LEP constraints} 
\label{subsec:LEP} 

The \texttt{HiggsBounds} program \cite{arXiv:0811.4169,arXiv:1102.1898,arXiv:1301.2345,arXiv:1311.0055} is a tool able to test a model against experimental 
data coming from LEP, Tevatron and the LHC. The program can be interfaced with \texttt{2HDMC} which will give appropriate inputs to \texttt{HiggsBounds}.
In our analysis we use \texttt{HiggsBounds} version 4.2.1 with the LEP experiment constraints only, in order to impose LHC constraints separately. 
\texttt{2HDMC} gives \texttt{HiggsBounds} a parton-level input for the three scalar Higgs bosons and the two charged ones. The exclusion test at $2\sigma$ 
is then performed on the five physical scalars of the theory. \texttt{HiggsBounds} returns a binary result indicating if the specific model point has been excluded 
at 95\% C.L. or not.

\subsection{LHC Higgs boson constraints}
\label{subsec:LHC}
What we call ``LHC constraints'' are restrictions coming from experimental results on the 125$~$GeV Higgs boson, \textit{i.e.} the 2HDM heavy Higgs boson $H$, in our case. To implement such limits, we use the exclusion contours in the plane of the signal strength for each individual production mode $\mu_{VBF/VH}$ vs $\mu_{ggh/tth}$ given by the combined ATLAS and CMS 
experiments at Run 1 \cite{Khachatryan:2016vau}. Assuming a Gaussian profile for the likelihood $\mathcal{L}$ at 68\% C.L., each exclusion contour for a specific decay channel 
$Y$ obeys  the following equation:
 \begin{align} \label{eq:ellipse}
  -2 \log \mathcal{L}_Y &\equiv \chi^2_Y  \\ \nonumber
	& = \left( \begin{array}{c}
		\mu_{ggH/ttH} - \widehat{\mu}_{ggH/ttH,Y}\\
		\mu_{VBF/VH} - \widehat{\mu}_{VBF/VH,Y}
                 \end{array} \right)^T
                 \left(\begin{array}{c c}
                        a_Y & b_Y \\
                        b_Y & c_Y
                 \end{array}\right)
                  \left( \begin{array}{c}
		\mu_{ggH/ttH} - \widehat{\mu}_{ggH/ttH,Y}\\
		\mu_{VBF/VH} - \widehat{\mu}_{VBF/VH,Y}
                 \end{array} \right) \,,
 \end{align}
where $\widehat{\mu}_{ggH/ttH,Y}$ and $\widehat{\mu}_{VBF/VH,Y}$ are the data best fit values and $a_Y$, $b_Y$ and $c_Y$ are the parameters of the ellipse. These five parameters fully describe the ellipse.
We fit the ellipses for each decay channel $Y=\{WW,ZZ,\gamma\gamma,\tau\tau,b\bar{b}\}$ and hence obtained the parametrisation for each of them (see \cite{Flament:2015wra} for more details).
We then compute the $\chi_Y^2$ value in the 2HDM using equation \ref{eq:ellipse}. For this, we assume the following relations:
\begin{equation}
 \left\{ \begin{array}{l}
      \mu_{ggH/ttH,Y} = \frac{(\sigma_{gg\rightarrow H}^{2HDM} + \sigma_{tt\rightarrow H}^{2HDM} )\times BR_Y^{2HDM}}{(\sigma_{gg\rightarrow H}^{SM} + \sigma_{tt\rightarrow H}^{SM})\times BR_Y^{SM}} \simeq \kappa_g^2\times \frac{BR_Y^{2HDM}}{BR_Y^{SM}} \vspace{0.5cm} \\
      \mu_{VBF/VH,Y} = \frac{(\sigma_{VBF}^{2HDM} + \sigma_{VH}^{2HDM} )\times BR_Y^{2HDM}}{(\sigma_{VBF}^{SM} + \sigma_{VH}^{SM})\times BR_Y^{SM}}  \simeq \kappa_V^2\times \frac{BR_Y^{2HDM}}{BR_Y^{SM}} \\
       \end{array}
 \right.
\end{equation}
with $\kappa_g^2\equiv \frac{\Gamma_{H\rightarrow gg}^{2HDM}}{\Gamma_{H\rightarrow gg}^{SM}}$, $\kappa_V^2\equiv \frac{\Gamma_{H\rightarrow WW}^{2HDM}}{\Gamma_{H\rightarrow WW}^{SM}} = \cos^2(\beta - \alpha)$ (see Table \ref{table:couplings}).
 
 Combining the log-likelihood ratios, we obtain:
 \begin{equation}
  \Delta \chi^2(p_j)=\sum_Y \chi^2_Y(p_j) - \sum_Y \chi^2_Y(\widehat{p_j})\,,
 \end{equation}
 with $p_j$ the set of free parameters on which the function depends and $\widehat{p_j}$ their value minimising the $\chi^2$ function. According to Wilks's theorem, the $\Delta \chi^2$ function follows a $\chi^2$ distribution with a number of degrees of freedom equal to the number of free parameters. In our case, we have six degrees of freedom ($\kappa_g^2$, $\kappa_V^2$, $BR_{H\rightarrow WW}^{2HDM}$, $BR_{H\rightarrow \tau\tau}^{2HDM}$, $BR_{H\rightarrow \gamma\gamma}^{2HDM}$, $BR_{H\rightarrow b\bar{b}}^{2HDM}$, as $BR_{H\rightarrow ZZ}^{2HDM}$ is linked to $BR_{H\rightarrow WW}^{2HDM}$). This choice in the free parameters implies that we assume there is no correlation between the kappas and the branching ratios, which is correct as long as the deviation of the branching ratio is not too large with respect to the Standard Model values.
A point in the 2HDM parameter space passing the LHC constraints, therefore, has a $\Delta\chi^2$ value lower than 12.85, which is the value 
at 95\% C.L. for a 6 degrees-of-freedom $\chi^2$ distribution.
 
\section{Search for a lighter scalar Higgs boson in the 2HDMs}
\label{sec:search}

A light resonance decaying into two photons is being searched for by CMS~\cite{CMS_diphoton} in the range of mass between 80 and 110 GeV. In this section we will explore the possibility that the signal may be given by the light scalar state in the 2HDMs. 
To compare with the experimental sensitivity (in particular at 8 TeV), we need to compute the expected production cross sections in the different production modes and 
branching ratios into the observed final states. In the following subsections we illustrate the procedure we followed and the results used in the present work to obtain
restrictions on the parameter space of the various 2HDMs. In \ref{subsec:XS_BR} we discuss the calculation method used for cross sections 
times branching ratios. Then we apply, in section \ref{subsec:const_free_param}, the present bounds coming from the three sets of constraints defined in the previous section in order to define the available parameter space. We finally test the sensitivity of the CMS low mass di-photon analysis at the LHC Run 1, in section \ref{subsec:CMS_analysis}, in the available parameter space for the four types of 2HDMs. To do so, we rely on a scan on the six free parameters in the physical basis.

\subsection{Cross sections and branching ratios}
\label{subsec:XS_BR}

We use the program \texttt{2HDMC} \cite{2HDMC} version 1.7.0 to compute the branching ratios of the different Higgs bosons of the theory. As input, the program requires a numerical value for each of the seven parameters of the physical basis and provides, as output, the total width, branching ratios and couplings at next to leading order (NLO) for each Higgs boson.

The cross sections can also be computed via programs like \texttt{SusHi} \cite{Harlander:2016hcx}: however, the output is restricted to the gluon fusion and $b \bar{b}$ production 
modes while \texttt{SusHi} does not provide vector boson fusion production (VBF) nor associated production with gauge bosons (VH). 
In order to overcome this restriction, and to quicken the calculation, we compute the cross sections using an approximation  that we have briefly introduced in section \ref{subsec:LHC} and that we denote in the following as the ``kappa trick''.
Defining the generic parameter $\kappa_Y$ as $\kappa_Y^2=\frac{\Gamma_Y^{2HDM}}{\Gamma_Y^{SM}}$ for a specific decay channel $Y$, we approximate the cross sections as:
\begin{equation}
 \sigma_{ggh}^{2HDM}\simeq \kappa_g^2\times\sigma_{ggh}^{SM}, \qquad \sigma_{VBF/VH}^{2HDM}\simeq \kappa_V^2\times\sigma_{VBF/VH}^{SM} = \sin^2(\beta-\alpha)\times\sigma_{VBF/VH}^{SM}\,.
\end{equation}
\label{eq:kappa_trick}
The second equation has such a simple form because, as the couplings of the light scalar Higgs boson to the W and Z bosons are rescaled in the same 
way compared to the SM couplings (cf Table \ref{table:couplings}), then $\kappa_Z=\kappa_W\equiv\kappa_V = \sin(\beta-\alpha)$.
The SM cross section is taken from the LHC Higgs Cross-Section Working Group \cite{Heinemeyer:2013tqa}. The kappas are computed thanks to the 
output given by \texttt{2HDMC}. Hence we are able to compute the cross section times branching ratio of the two neutral scalar Higgs bosons using only 
the 2HMDC program, via equation \ref{eq:kappa_trick}.

It is pertinent at this point to comment on the level of validity of this approximation. The cross section production in VBF and VH mode 
should not cause any problem as the leading effect arises at tree level, however for the gluon fusion mode a loop induced coupling is present and thus it is important to check the validity 
of the ``kappa trick'' for this production mode. Note indeed that for loop induced vertices the use of an effective kappa factor is not always appropriate
and more general parameterisations exist (see for example \cite{Cacciapaglia:2009ky,Cacciapaglia:2014rla}).
In order to explore this issue and establish if this simple approximation could be used, we performe a comparison between the cross sections in gluon 
fusion obtained via the program \texttt{SusHi} and the ones obtained with the ``kappa trick''.
As \texttt{2HDMC} only considers NLO corrections, we also ran \texttt{SusHi} at NLO. The SM inputs required by the two programs are set to the recommended 
values given by the Particle Data Group \cite{PDG} summarised in Table \ref{table:SM_inputs}. The 2HDM inputs used are given in Table \ref{table:param_comparison}: we chose to fix all the parameters except the mass of the light neutral scalars, whose cross section we want to test.
We use \texttt{SusHi} version 1.6.0 together with \texttt{LHAPDF} 6.1.6 \cite{Buckley:2014ana}. The parton distribution functions used in the program are \texttt{MMHT201468cl} 
for LO and \texttt{PDF4LHC15$\_$mc} for NLO and NNLO \cite{Butterworth:2015oua}. The renormalization and factorization scales $\mu_R$ and $\mu_F$ for the gluon 
fusion process are set to $\mu_R=\mu_F=m_{\phi}/2$ with $\phi=\{h,H,A\}$ \cite{Harlander:2013qxa}. The $b\bar{b}$ production mode proposed by \texttt{SusHi} is turned off.

\begin{table}[h!]
 \begin{center}
  \begin{tabular}{c|c|c|c|c|c}
  $m_W$ (GeV) & $\Gamma_W$ (GeV) & $m_Z$ (GeV) & $\Gamma_Z$ (GeV) & $\bar{m_b}(m_b)$ (GeV) & $m_t$(pole) (GeV)  \\ \hline
   80.385 & 2.085 & 91.1876 & 2.4952 & 4.18 & 173.34 \\ \hline
  \end{tabular}\vspace{0.5cm}
  \begin{tabular}{c|c|c|c|c}
  $m_c$(pole) (GeV)  & $\alpha_{EM}$ &$\alpha$ & $\alpha_s$ & $G_F$ (GeV)$^{-2}$ \\ \hline
  1.76 & 1/127.934 & 1/137.0359991 & 0.118 & 1.16637$\times 10^{-5}$ \\ \hline
  \end{tabular}
 \end{center}
 \caption{SM input parameters \cite{PDG}.}
 \label{table:SM_inputs}
\end{table}

\begin{figure}[h!]
 \centering
\includegraphics[width=0.7\textwidth]{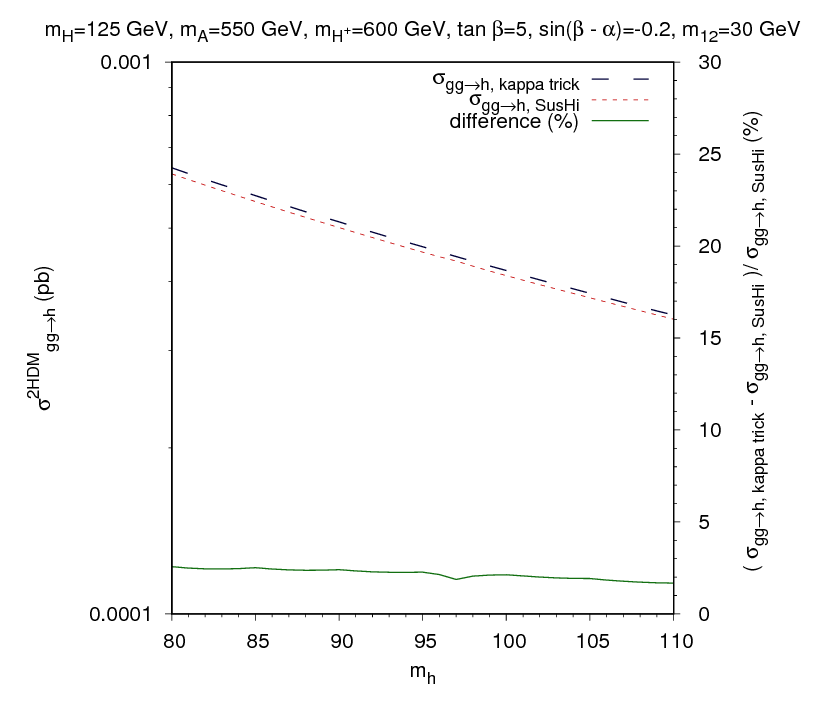}
\caption{$\sigma^{2HDM}_{gg\rightarrow h}$  computed with the ``kappa trick'' (dashed blue line), with \texttt{SusHi} (dotted red line) and the deviation between the two (solid green line).}
\label{fig:comparison_sushi}
\end{figure}

Fixing six of the seven free parameters, we allow only $m_h$ to vary between 80 to 110 GeV with a step of 1 GeV between each point (see 
Table \ref{table:param_comparison}). The results are plotted in Figure~\ref{fig:comparison_sushi}. The dashed blue line corresponds to the cross section 
computed with the ``kappa trick'', the dotted red line to the computation with \texttt{SusHi} and the solid green line to the deviation between the two, computed as: 
 \begin{equation}
  \Delta \equiv \frac{\sigma_{gg\rightarrow h}^{kappa~ trick}  - \sigma_{gg\rightarrow h}^{\texttt{SusHi}}}{\sigma_{gg\rightarrow h}^{\texttt{SusHi}} }\times 100\,.
 \end{equation}
 
\begin{table}[h!]
 \begin{center}
  \begin{tabular}{c|c|c|c|c|c|c}
   $m_h$ (GeV) & $m_H$ (GeV) & $m_A$ (GeV) & $m_{H^{\pm}}$ (GeV) & $\tan\beta$ & $\sin(\beta-\alpha)$ & $m_{12}$ (GeV)  \\ \hline
   [80;110], step 1. & 125 & 550 & 600 & 5 & -0.2 & 30 \\ \hline
  \end{tabular}
 \end{center}
 \caption{Input parameters in the 2HDM Type I for the comparison between \texttt{SusHi} and ``kappa trick'' cross sections.}
 \label{table:param_comparison}
\end{table}

The plot shows a deviation of less than  3\% for the whole mass range  and this deviation is stable upon modification of the 
values of the input parameters (see Figure~\ref{fig:comparison_sushi_mA_80} in the appendix). As it stays within the range allowed by the uncertainties (theoretical, PDF and $\alpha_s$) calculated by the LHC Higgs Cross-Section Working Group \cite{Heinemeyer:2013tqa}, we consider this test as having validated our method for the light Higgs boson.
For completeness, we make a similar analysis for the heavy Higgs boson at 125 GeV (see Figure~\ref{fig:XS_vs_mH} in the appendix) finding deviations less 
than 1\% at $m_H=125$ GeV.
In the rest of the study, therefore, we will use the ``kappa trick'' approximation to compute the cross section of the light and heavy scalar Higgs bosons in 
the 2HDMs.

 \subsection{Constraining the 2HDMs parameter space}
 \label{subsec:const_free_param}
 
In this section we study the influence of the three sets of constraints defined in Section \ref{sec:bounds} (indirect, LEP and LHC constraints) on the free parameters.
For this purpose we generate a set of one million points for each of the four different types of model defined in Table \ref{table:types} with random values for each of the 
free parameters. The available ranges we use in the simulation are given in Table \ref{table:input_param}. The range of variation for $m_h$ corresponds to the mass range available in the CMS di-photon analysis. The lower bound of 80 GeV for $m_{H^{\pm}}$ comes from the bound obtained at the LEP experiment \cite{Abbiendi:2013hk}. The ranges for $m_A$ and $m_{12}^2$, although not totally general, are the result of previous quick scans that we will not show in this paper and which eliminate areas with a very low density of points passing the three sets of constraints (indirect, LEP and LHC constraints).

Once the points are generated, we impose the three kinds of constraints detailed above: indirect ones, direct LEP and LHC Higgs boson ones. 

\begin{table}[h!]
  \begin{center}
    \begin{tabular}{c|c|c|c|c|c|c}
      $m_h$ (GeV) & $m_H$ (GeV) & $m_A$ (GeV) & $m_{H^{\pm}}$ (GeV) & $\sin(\beta-\alpha)$ & $\tan\beta$ & $m_{12}^2$ (GeV)$^2$ \\ \hline
      [80;110] & 125 & [60;1000] & [80;1000] & [-1;1] & [1/50;50] & [-(300)$^2$;+(200)$^2$]
    \end{tabular}
  \end{center}
  \caption{Range of variation for the free parameters used in the analysis.}
  \label{table:input_param}
\end{table}

  \begin{figure}[h!]
 \begin{subfigure}{0.48\textwidth}
  \centering
  \includegraphics[width=\textwidth]{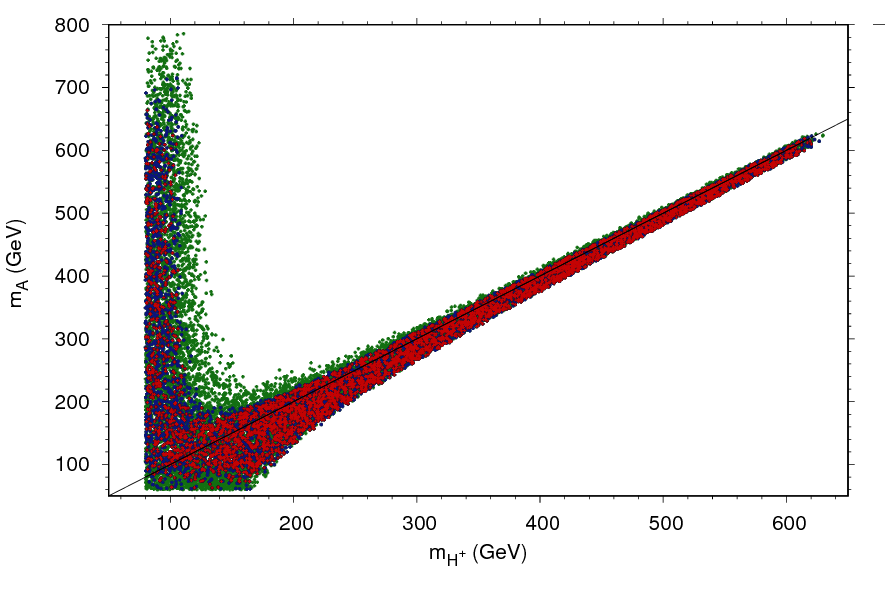}
  \end{subfigure}
  \hfill
  \begin{subfigure}{0.48\textwidth}
  \includegraphics[width=\textwidth]{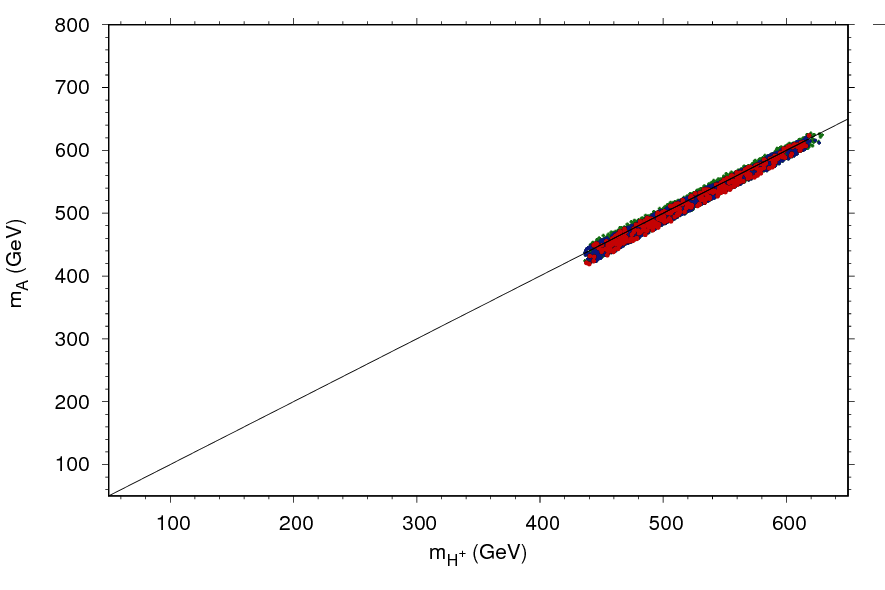}
  \end{subfigure}\\
  \begin{subfigure}{0.48\textwidth}
  \includegraphics[width=\textwidth]{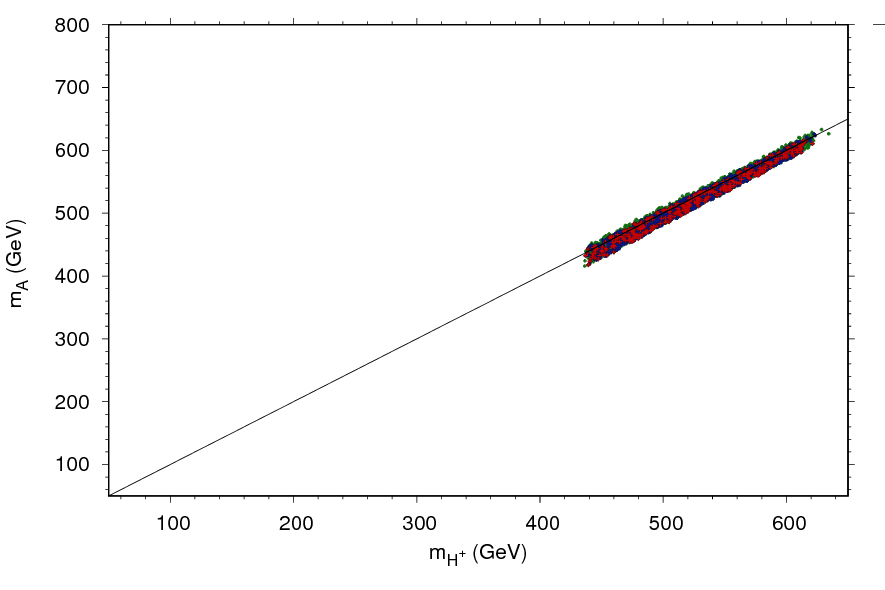}
  \end{subfigure}
  \hfill
  \begin{subfigure}{0.48\textwidth}
  \includegraphics[width=\textwidth]{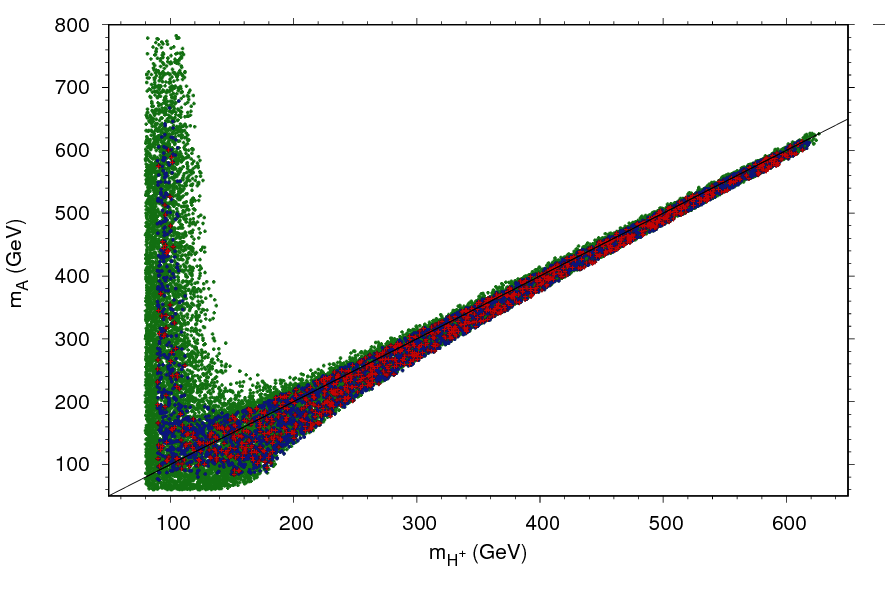}
  \end{subfigure}
  \caption{Constraints on the free parameters in the plane $m_A$ vs $m_{H^{\pm}}$. Top left: Type I. Top right: Type II. Bottom left: Flipped. Bottom right: Lepton Specific. In green: points passing indirect constraints only. In blue: points passing indirect and LEP constraints. In red: points passing indirect, LEP and LHC constraints.}
  \label{fig:mA_vs_mHpm}
 \end{figure}
 
In Figure~\ref{fig:mA_vs_mHpm}, all the generated points are plotted in the plane $m_A$ vs $m_{H^{\pm}}$. The upper left panel corresponds to Type I, the upper 
right to Type II, the lower left to the Flipped model and the lower right to Lepton Specific model. The points passing only the indirect constraints are plotted in green, 
those passing indirect and LEP constraints are in blue and those passing indirect, LEP and LHC constraints are in red. We will use these same conventions in the rest of this section.
 
Firstly we can see that the $m_A$ and $m_{H^{\pm}}$ masses are very correlated: when $m_A$ and $m_{H^{\pm}}$ grow, the indirect constraints force them to be near the black line corresponding to $m_A=m_{H^{\pm}}$. This is due to the T parameter which is very sensitive to these two masses and enforces them to be close to each other.
Looking only at the red points, those which pass the three sets of constraints we defined previously, we can see that the two masses are bounded. In Type I, 
we find that most of the red points lie in the ranges $m_A\in$[60~GeV;~650~GeV] and $m_{H^{\pm}}\in$[80~GeV;~630~GeV]. In Type II and Flipped, the two masses are much more constrained $m_A\in$[400 GeV; 650 GeV] and $m_{H^{\pm}}\in$[430 GeV; 630 GeV]: this is due to the fact that the down-type quarks couple now to the $\phi_1$ doublet instead of the $\phi_2$ doublet as in Type I, thus the $\mathcal{BR} (B\rightarrow X_s\gamma)$ flavour limit imposes a very strong constraint on the mass of the charged Higgs bosons (see Figure~\ref{fig:BXS_lim_Type2} in the appendix). Associated with the T parameter constraint, it imposes also the bounds on the pseudo-scalar mass.
 The Lepton Specific case is very similar to Type I as the couplings of the down-type quark are the same. We find $m_A\in$[80~GeV;~630~GeV] and 
 $m_{H^{\pm}}\in$[80 GeV; 630 GeV] to be the preferred regions. We should remark that these bounds are not absolute and that there may be red points exceeding these bounds. However, our simulation shows that the bulk of the allowed points are inside the ranges, so that we decided to use them in order to increase the statistics of our scan.
 
 \begin{figure}[h!]
 \begin{subfigure}{0.48\textwidth}
  \centering
  \includegraphics[width=\textwidth]{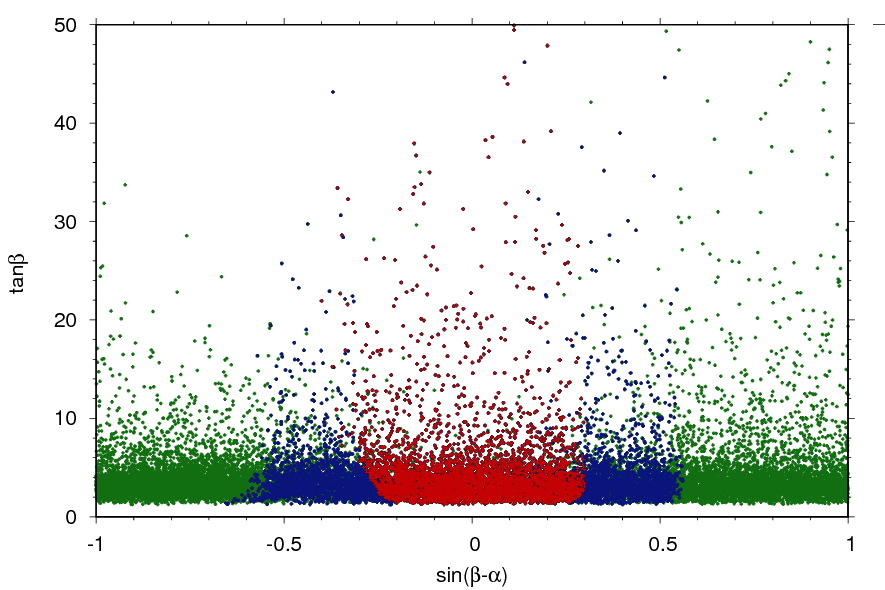}
  \end{subfigure}
  \hfill
  \begin{subfigure}{0.48\textwidth}
  \includegraphics[width=\textwidth]{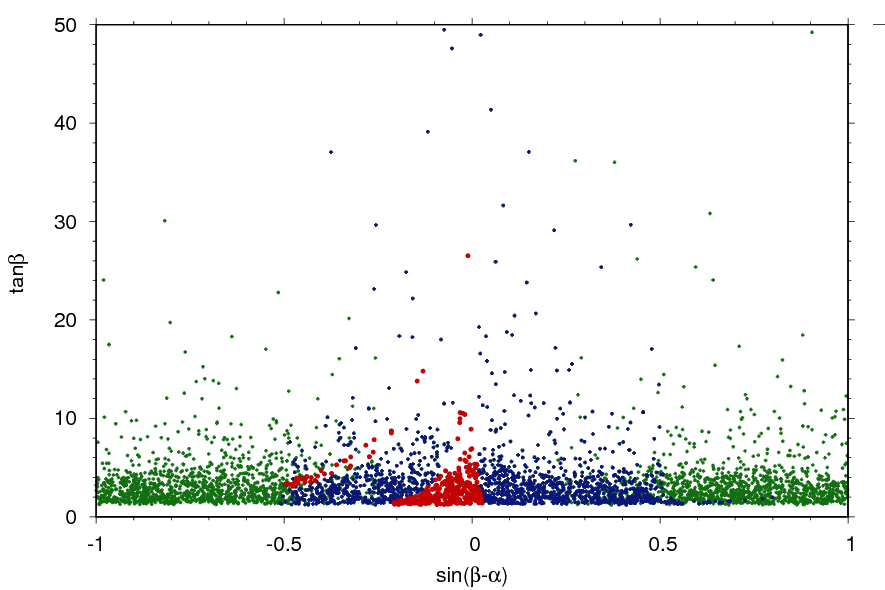}
  \end{subfigure}\\
    \begin{subfigure}{0.48\textwidth}
    \includegraphics[width=\textwidth]{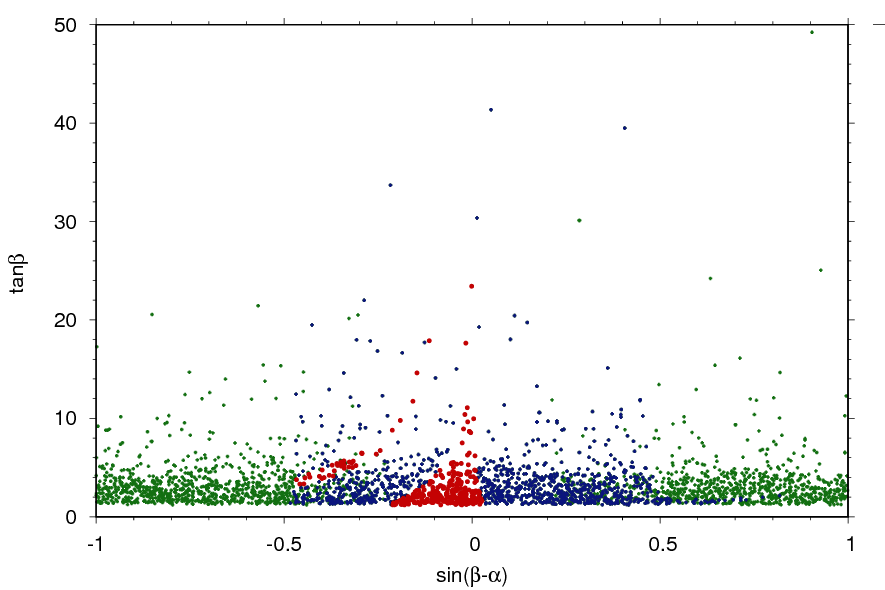}
  \end{subfigure}
  \hfill
  \begin{subfigure}{0.48\textwidth}
  \includegraphics[width=\textwidth]{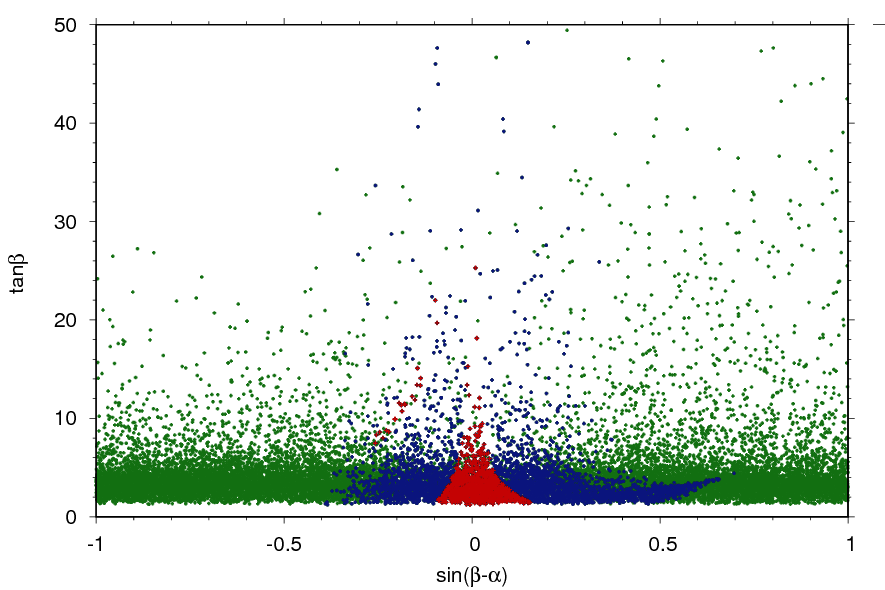}
  \end{subfigure}
  \caption{Constraints on the free parameters in the plane $\tan\beta$ vs $\sin(\beta-\alpha)$. Top left: Type I. Top right: Type II. Bottom left: Flipped. Bottom right: Lepton Specific. Same colour code as in Figure~\ref{fig:mA_vs_mHpm}.}
  \label{fig:tanB_vs_sinBA}
 \end{figure}

Looking now at the plane $\tan\beta$ vs $\sin(\beta-\alpha)$ (shown in Figure~\ref{fig:tanB_vs_sinBA}) we can constrain in the same way the $\tan\beta$ parameter. 
If it is difficult to impose an upper limit in all types as we lack statistics for high values of $\tan\beta$ and we see a few red points up to the upper value, nevertheless we can impose a lower bound of 
$\tan\beta>1.2$ for the four different types.

  \begin{figure}[h!]
 \begin{subfigure}{0.48\textwidth}
  \centering
  \includegraphics[width=\textwidth]{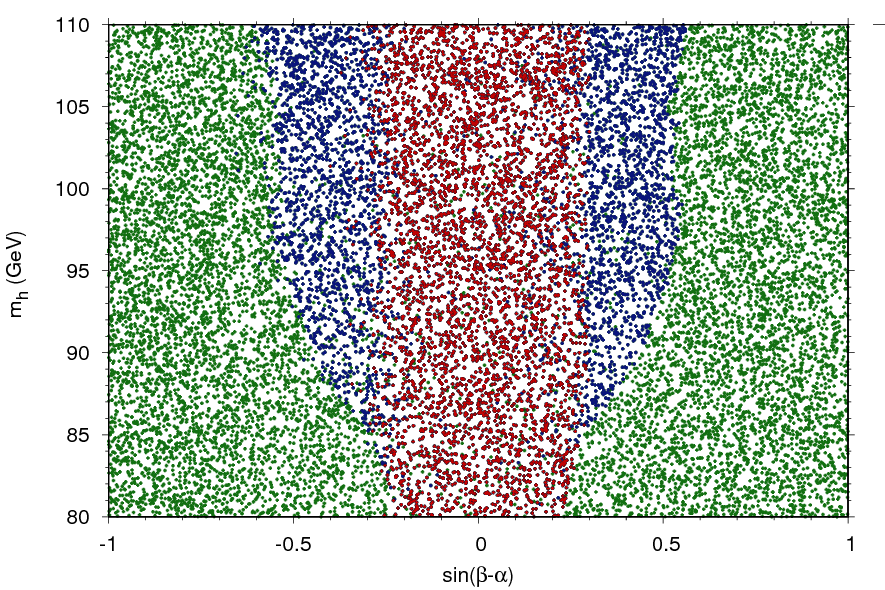}
  \end{subfigure}
  \hfill
  \begin{subfigure}{0.48\textwidth}
   \includegraphics[width=\textwidth]{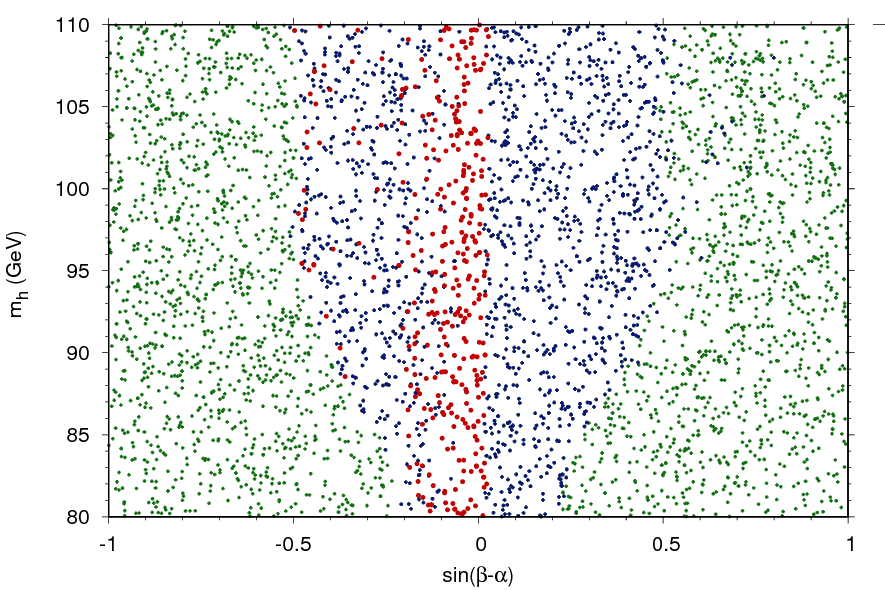}
   \end{subfigure}
    \begin{subfigure}{0.48\textwidth}
  \centering
  \includegraphics[width=\textwidth]{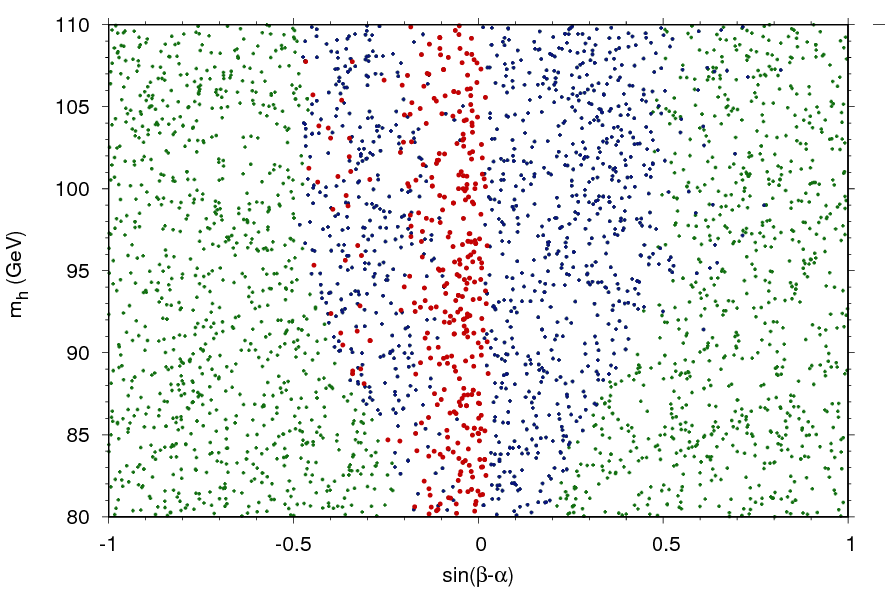}
  \end{subfigure}
  \hfill
  \begin{subfigure}{0.48\textwidth}
   \includegraphics[width=\textwidth]{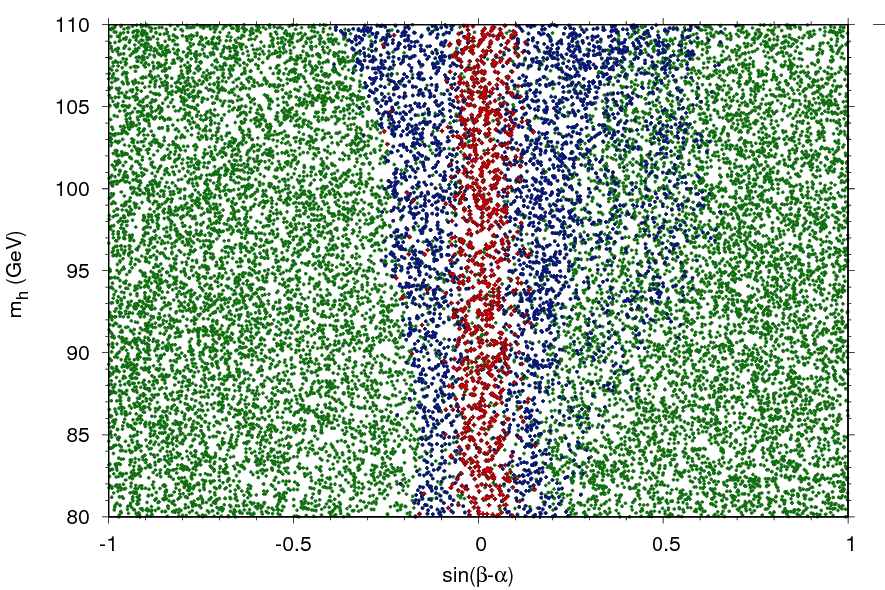}
   \end{subfigure}
  \caption{Constraints on the free parameters in the plane $m_h$ vs $\sin(\beta-\alpha)$. Top left: Type I. Top right: Type II. Bottom left: Flipped. Bottom right: Lepton Specific. Same colour code as in Figure~\ref{fig:mA_vs_mHpm}.}
  \label{fig:mh_vs_sinBA}
 \end{figure}
 
 The bounds on $\sin(\beta - \alpha)$ can be more easily seen in the plane $m_h$ vs $\sin(\beta-\alpha)$ (shown in Figure~\ref{fig:mh_vs_sinBA}): we see that $m_h$ is not constrained as red points span the whole range of masses. For $\sin(\beta - \alpha)$, the allowed range is close to zero, which is consistent with our choice of $m_H = 125$ GeV: as $\sin(\beta-\alpha) \simeq 0$, we have $\cos(\beta-\alpha) \simeq 1$ which means that the couplings of the heavy Higgs boson $H$ to the gauge bosons are close to the SM ones. We are therefore close to the alignment limit \cite{Bernon:2015wef}.
 We find that the preferred ranges are $\sin(\beta-\alpha)\in$[-0.4; 0.3] for Type I, $\sin(\beta-\alpha)\in$[-0.5; 0.05] for Type II and Flipped model and $\sin(\beta-\alpha)\in$[-0.3; 0.2] for Lepton Specific model.
 
 \begin{figure}[h]
 \begin{subfigure}{0.4\textwidth}
  \centering
  \includegraphics[width=\textwidth]{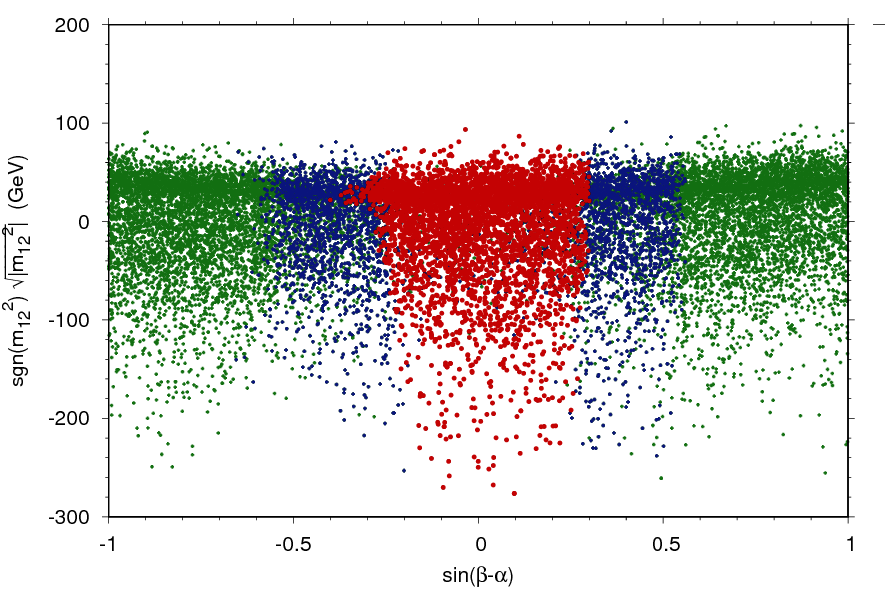}
  \end{subfigure}
  \hfill
  \begin{subfigure}{0.4\textwidth}
   \includegraphics[width=\textwidth]{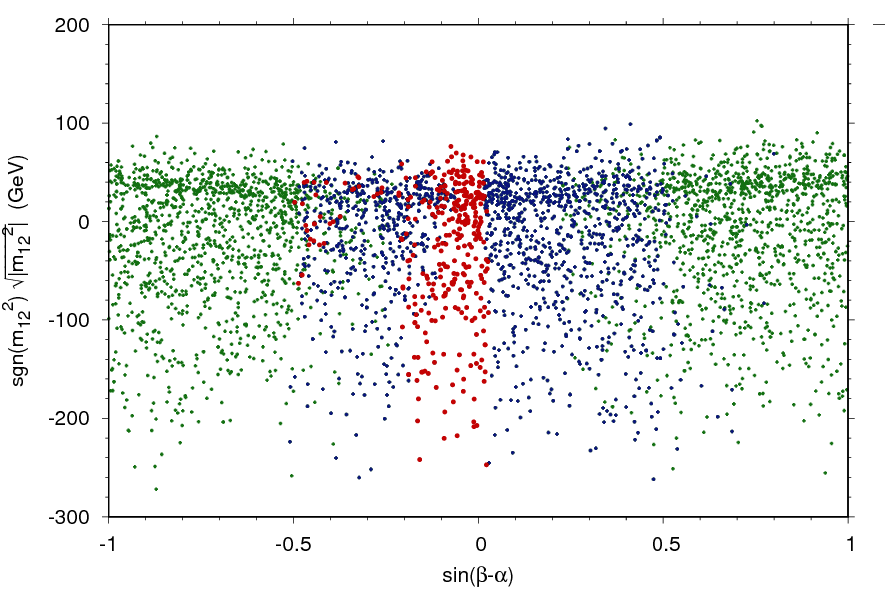}
   \end{subfigure}
    \begin{subfigure}{0.4\textwidth}
  \centering
  \includegraphics[width=\textwidth]{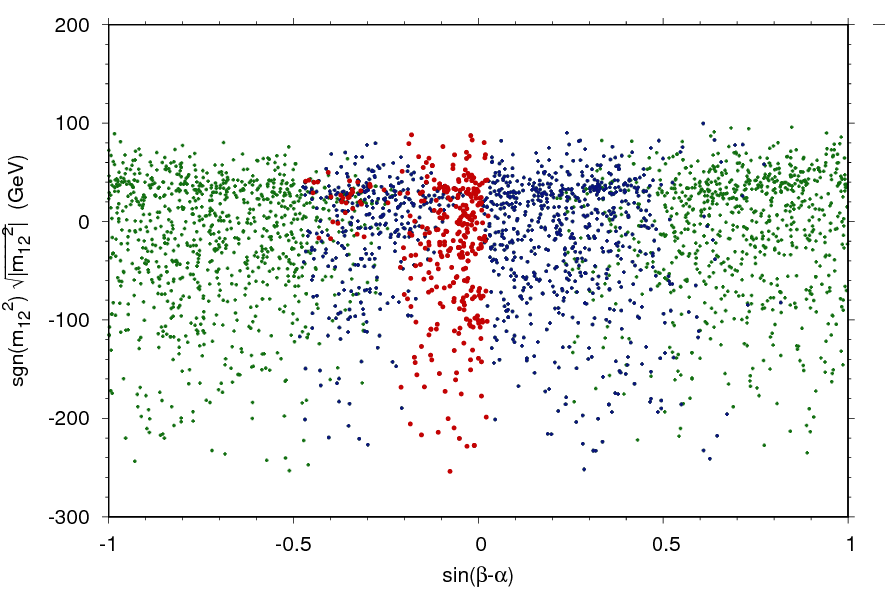}
  \end{subfigure}
  \hfill
  \begin{subfigure}{0.4\textwidth}
   \includegraphics[width=\textwidth]{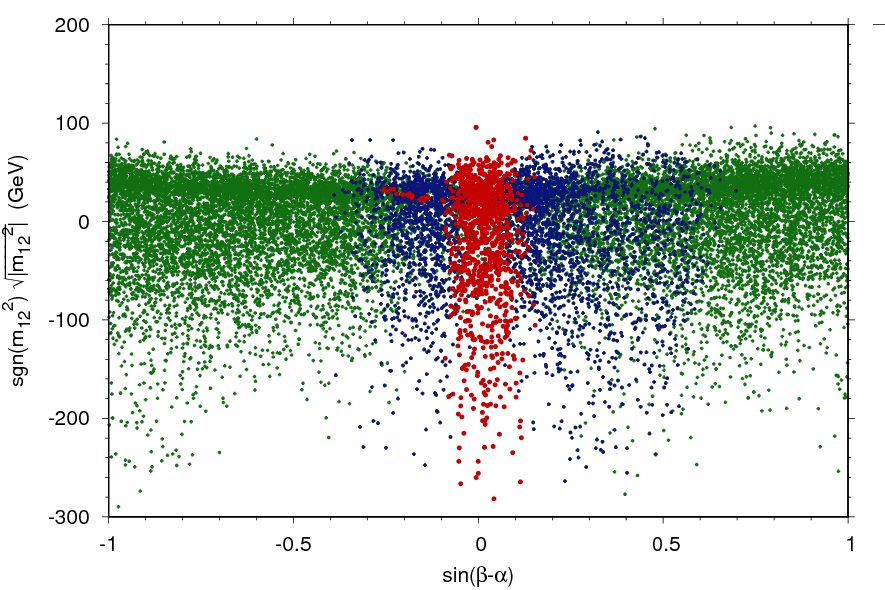}
   \end{subfigure}
  \caption{Constraints on the free parameters in the plane $m_{12}$ vs $\sin(\beta-\alpha)$. Top left: Type I. Top right: Type II. Bottom left: Flipped. Bottom right: Lepton Specific.  Same colour code as in Figure~\ref{fig:mA_vs_mHpm}.}
  \label{fig:m12_vs_sinBA}
 \end{figure}
 
Finally, looking at the plane $m_{12}$ vs $\sin(\beta-\alpha)$ we can constrain the last free parameter (see Figure~\ref{fig:m12_vs_sinBA}). We cannot put any lower bound on $m_{12}^2$ but we find $m_{12}^2<$(100 GeV)$^2$ in the four different types.
 
\begin{figure}[h!]
 \begin{subfigure}{0.4\textwidth}
  \centering
  \includegraphics[width=\textwidth]{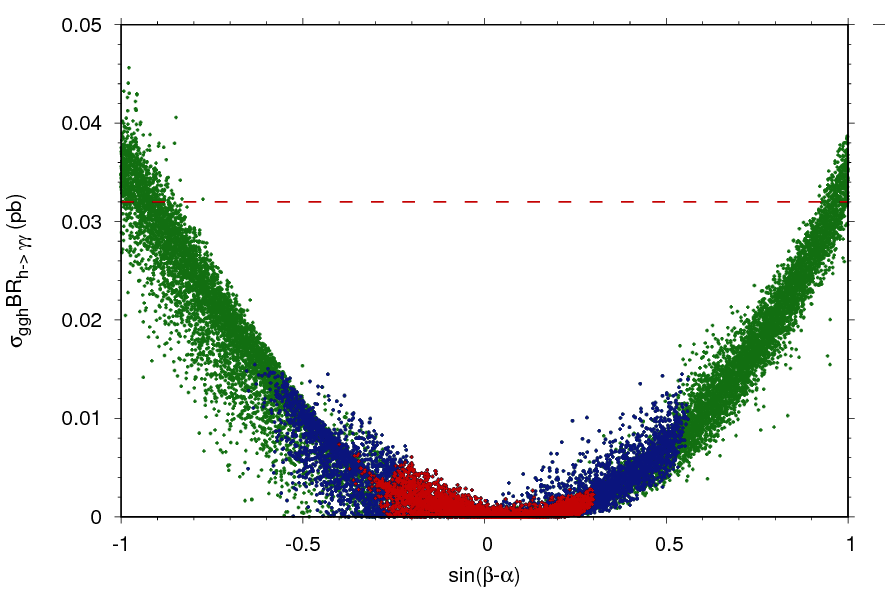}
  \end{subfigure}
  \hfill
  \begin{subfigure}{0.4\textwidth}
   \includegraphics[width=\textwidth]{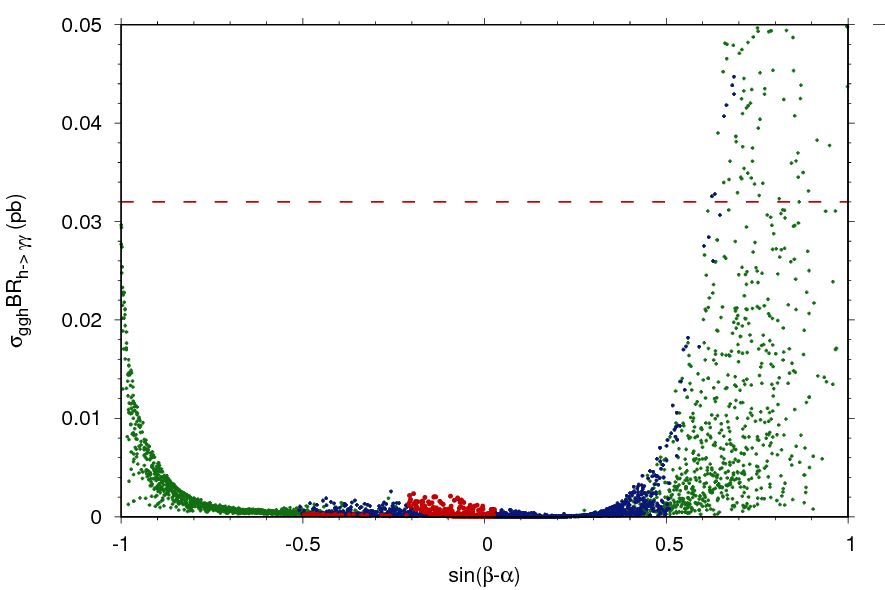}
   \end{subfigure}\\
 \begin{subfigure}{0.4\textwidth}
  \centering
  \includegraphics[width=\textwidth]{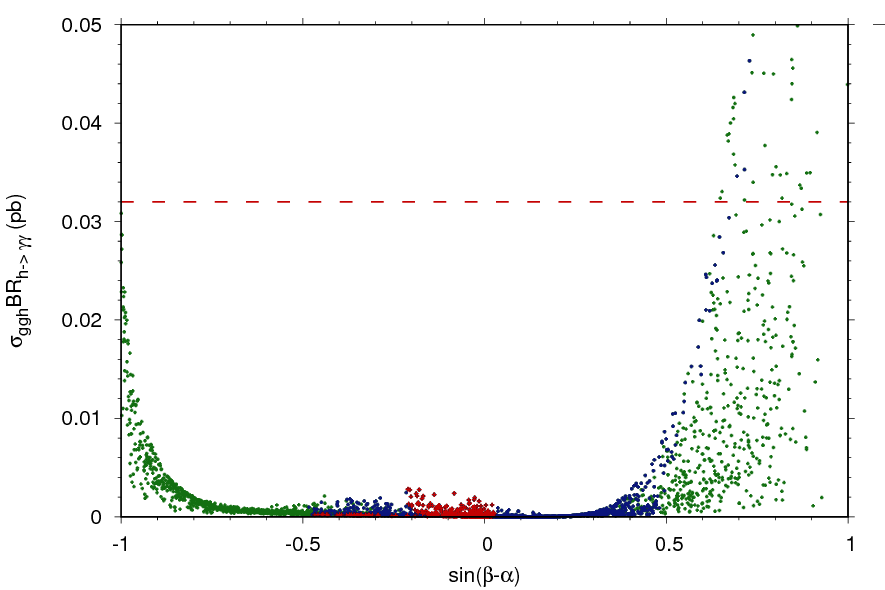}
  \end{subfigure}
  \hfill
  \begin{subfigure}{0.4\textwidth}
   \includegraphics[width=\textwidth]{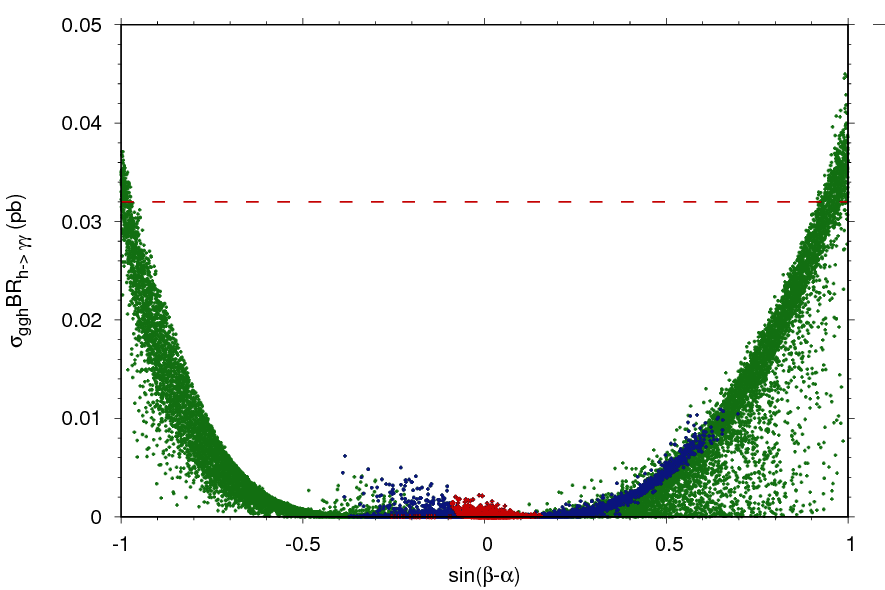}
   \end{subfigure}
   \caption{2HDM generated points in the plane $\sigma \times BR_{h\rightarrow \gamma\gamma}$ vs $sin(\beta-\alpha)$ in the gluon fusion production mode. Top left: Type I. Top right: Type II. Bottom left: Flipped. Bottom right: Lepton Specific. Same colour code as in Figure~\ref{fig:mA_vs_mHpm}. The dashed line corresponds to the minimum value of the CMS observed upper limit in the gluon fusion production mode.}
  \label{fig:XS_BR_vs_sinBA_ggh}
 \end{figure}

\begin{figure}[h!]
   \begin{subfigure}{0.48\textwidth}
  \centering
  \includegraphics[width=\textwidth]{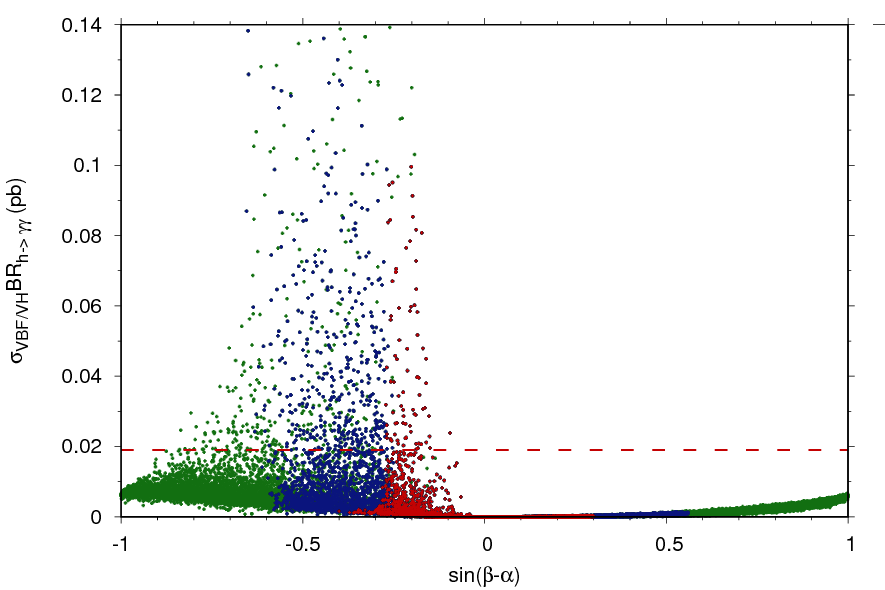}
  \end{subfigure}
  \hfill
  \begin{subfigure}{0.48\textwidth}
   \includegraphics[width=\textwidth]{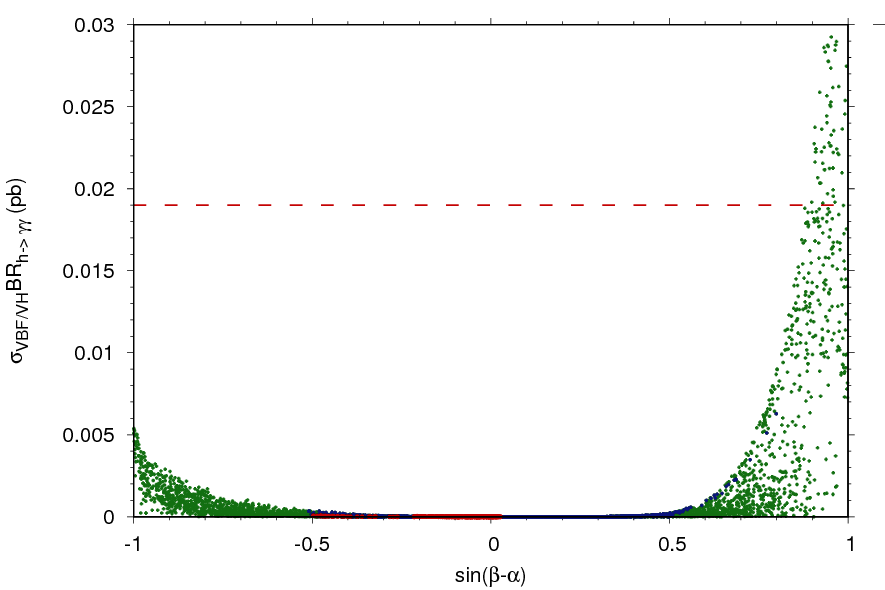}
   \end{subfigure}\\
      \begin{subfigure}{0.48\textwidth}
  \centering
  \includegraphics[width=\textwidth]{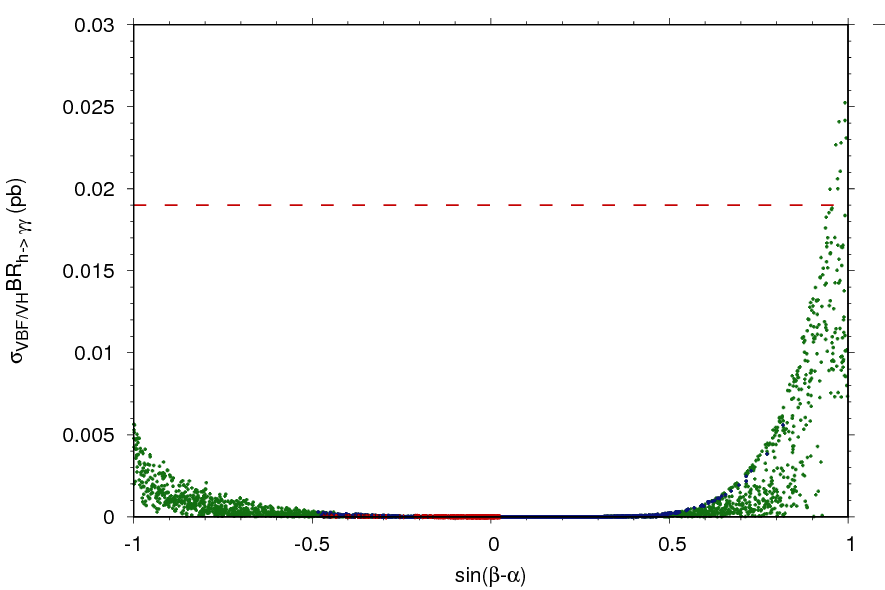}
  \end{subfigure}
  \hfill
  \begin{subfigure}{0.48\textwidth}
   \includegraphics[width=\textwidth]{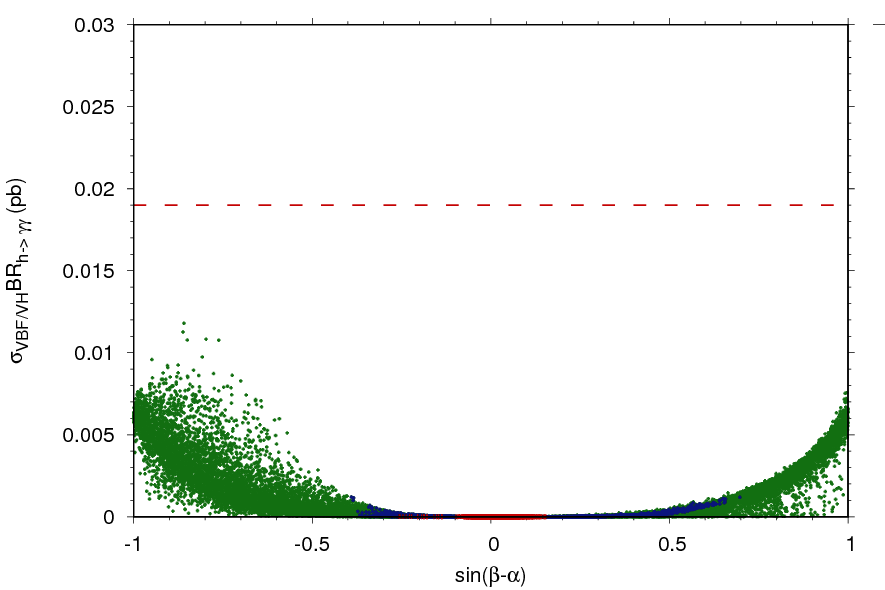}
   \end{subfigure}
   \caption{2HDM generated points in the plane $\sigma \times BR_{h\rightarrow \gamma\gamma}$ vs $\sin(\beta-\alpha)$ in the VBF/VH production mode. Top left: Type I. Top right: Type II. Bottom left: Flipped. Bottom right: Lepton Specific. Same colour code as in Figure~\ref{fig:mA_vs_mHpm}. The dashed line corresponds to the minimum value of the CMS observed upper limit in the VBF/VH production mode.}
  \label{fig:XS_BR_vs_sinBA_VBFVH}
 \end{figure}

 The previous results show that the range of the free parameters can be further limited in order to increase the statistics of the allowed points. 
In addition to this, as we are interested in checking the sensitivity to a lighter Higgs boson at LHC Run$~$1 in the di-photon decay channel, we can further restrict the areas of 
interest to where the red points correspond to relatively high values of cross section times branching ratio to two photons. The minimum value of the CMS observed upper limit \cite{CMS_diphoton} is 0.032 pb in the gluon fusion channel, obtained for $m_h=103$ GeV and 0.019 pb in the VBF/VH channel, obtained for $m_h=100.5$ GeV. Keeping these values in mind, we can look at the predicted 2HDM cross-section times branching ratio values as a function of 
$\sin(\beta - \alpha)$. We plot the results for the gluon fusion production mode in Figure~\ref{fig:XS_BR_vs_sinBA_ggh} and for VBF/VH production mode in 
Figure~\ref{fig:XS_BR_vs_sinBA_VBFVH}. 
The red dotted line corresponds to the minimum value of the CMS observed upper limit for each of the production modes. If all the red points 
are below this line, it means that CMS was not sensitive to a lighter Higgs boson in this particular channel at LHC Run 1. 
 
The first important result we can extract from these figures is that in the Type II, Flipped and Lepton Specific models, CMS had no sensitivity to a lighter Higgs 
boson at LHC Run 1 in the $h \rightarrow \gamma \gamma$ decay channel, neither in the gluon fusion nor in the VBF/VH production mode. Therefore we will not carry on with these types any further.
Looking at the results for Type I, we can see that there is no sensitivity in the gluon fusion channel. However, in the VBF/VH channel, we find red points above the 
dashed line. As the value of the CMS observed upper limit depends on the mass of the light Higgs boson considered, the dashed line represented on the plots 
is not an absolute bound. Some of the red points above it can be \textit{de facto} below the CMS observed limit, but it is a good indication of the potential capability of the channel for some exclusion. We can therefore expect to have some sensitivity in the VBF/VH channel.
  
We can exploit Figure~\ref{fig:XS_BR_vs_sinBA_VBFVH} even further by choosing to look only at areas where the points have relatively high values of cross section times branching ratio, i.e. areas where the points are close to the CMS analysis limit sensitivity. We choose a lower bound 
at 0.01 pb to select the points, which corresponds to $\sin(\beta - \alpha)\in$[-0.3;-0.05].
\begin{figure}[h!]
 \begin{subfigure}{0.48\textwidth}
  \centering
  \includegraphics[width=\textwidth]{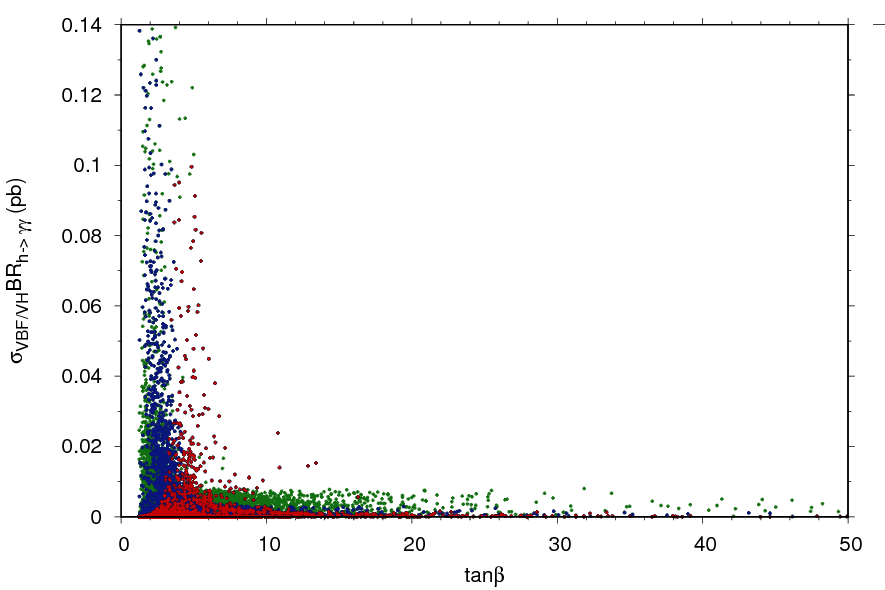}
  \end{subfigure}
  \hfill
  \begin{subfigure}{0.48\textwidth}
   \includegraphics[width=\textwidth]{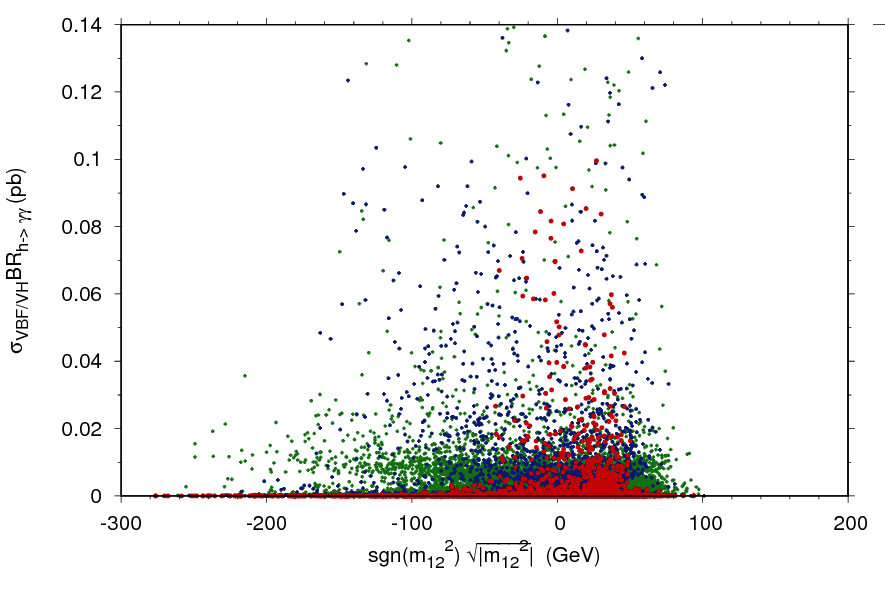}
   \end{subfigure}
  \caption{Value of the cross section times branching ratio in the VBF/VH production mode as a function of $\tan\beta$ (left) and $sgn(m_{12})\times\sqrt{|m_{12}^2|}$ (right) in Type I. Same colour code as in Figure~\ref{fig:mA_vs_mHpm}.}
  \label{fig:XS_BR_vs_tanB_m12}
 \end{figure}
 
We can similarly work with tighter ranges for the parameters $\tan\beta$ and $m_{12}^2$ (see Figure~\ref{fig:XS_BR_vs_tanB_m12}). We choose $\tan\beta\in$[2;12] 
and $m_{12}^2\in$[-(100 GeV)$^2$;+(100 GeV)$^2$].

\clearpage
After having defined the allowed parameter region, and the more promising region with respect to the di-photon search, we are now ready to perform a 
second ``focused'' simulation and make a detailed comparison with the sensitivity of the CMS search at 8 TeV.

\subsection{Comparison with the CMS low mass di-photon analysis}
 \label{subsec:CMS_analysis}
 
We thus perform a new scan with one million points, this time for Type I only, using the restricted parameter ranges we found in the previous section (see Table \ref{table:input_param2}). 
We remind the reader that for $m_A$ and $m_{H^{\pm}}$ the new range results only from the three sets of constraints (the indirect, LEP and LHC constraints) we imposed. For the parameters $\sin(\beta-\alpha)$, $\tan\beta$ 
and $m_{12}^2$ it results from our choice to restrict the scan to areas with large value of $\sigma_{VBF/VH}\times$BR$_{h\rightarrow \gamma \gamma}$ 
(above 0.01 pb), as explained in Section \ref{subsec:const_free_param}.
\begin{table}[h!]
  \begin{center}
    \begin{tabular}{c|c|c|c|c|c|c}
      $m_h$ (GeV) & $m_H$ (GeV) & $m_A$ (GeV) & $m_{H^{\pm}}$ (GeV) & $\sin(\beta-\alpha)$ & $\tan\beta$ & $m{12}^2$ \\ \hline
      [80;110] & 125 & [60;650] & [80;630] & [-0.3;-0.05] & [2;12] & [-(100)$^2$;+(100)$^2$]
    \end{tabular}
  \end{center}
  \caption{Allowed range of variation for the free parameters.}
  \label{table:input_param2}
\end{table}

\begin{figure}[h!]
    \centering
    \begin{subfigure}[b]{.48\textwidth}
      \includegraphics[width=\textwidth]{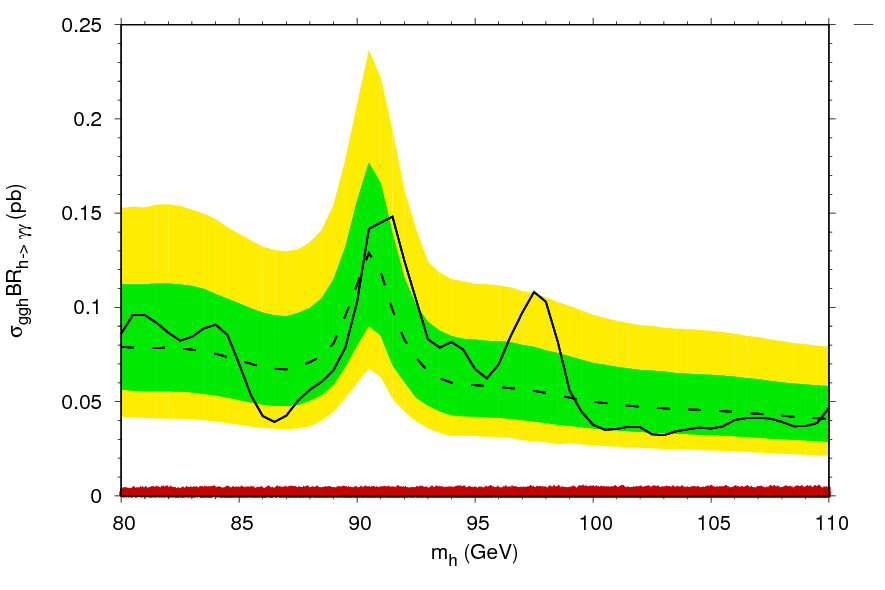}
    \end{subfigure}
    \hfill
    \begin{subfigure}[b]{.48\textwidth}
      \includegraphics[width=\textwidth]{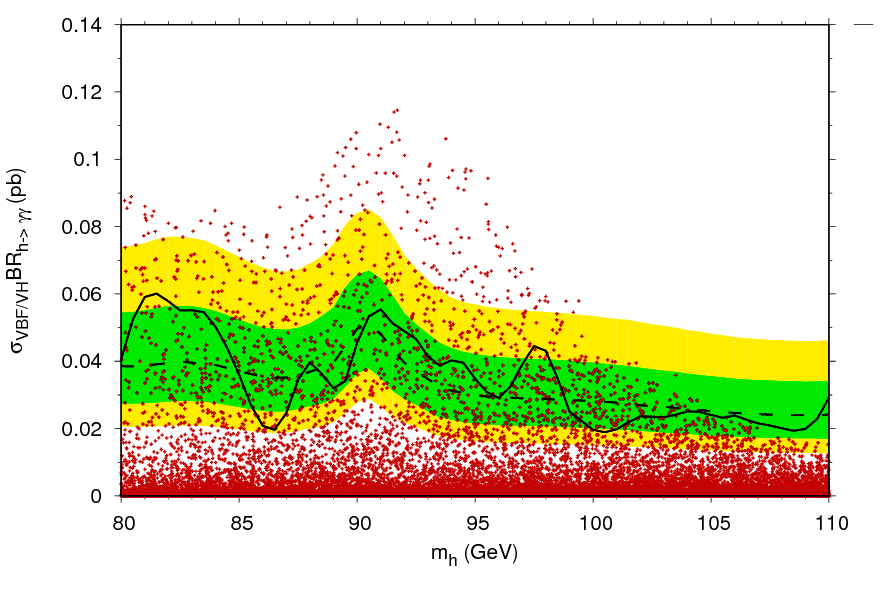}
    \end{subfigure}    
      \caption{Points generated in the 2HDM Type I passing indirect, LEP and LHC constraints, superimposed on the results of the CMS 8 TeV low-mass di-photon analysis \cite{CMS_diphoton} in the  gluon fusion production mode (left panel) and the combined VBF and VH production mode (right panel). The dashed line corresponds to the expected upper limit on $\sigma\times BR_{h\rightarrow \gamma\gamma}$ at 95\% C.L., with 1 and 2 sigma errors in green and yellow respectively. The solid line is the observed upper limit at 95\% C.L.}
      \label{fig:XS_BR_CMS}
  \end{figure}

The resulting points of this second scan are plotted in Figure~\ref{fig:XS_BR_CMS} in the plane $\sigma\times~BR_{h\rightarrow \gamma\gamma}$ in the gluon fusion production mode (left panel) and the VBF/VH production mode (right panel) vs $m_h$, superimposed on the public exclusion limits of CMS collaboration. For convenience only the red points, \textit{i.e.} the points passing all of the indirect, LEP and LHC constraints, are 
plotted here. 
The results confirm our expectation from Figures \ref{fig:XS_BR_vs_sinBA_ggh} and \ref{fig:XS_BR_vs_sinBA_VBFVH} that there is no sensitivity in the gluon 
fusion production mode but many points are above the CMS observed limit in the VBF/VH production mode for a light Higgs boson with mass below 105 GeV.
  
\begin{figure}[h!]
    \centering
    \begin{subfigure}[b]{.48\textwidth}
      \includegraphics[width=\textwidth]{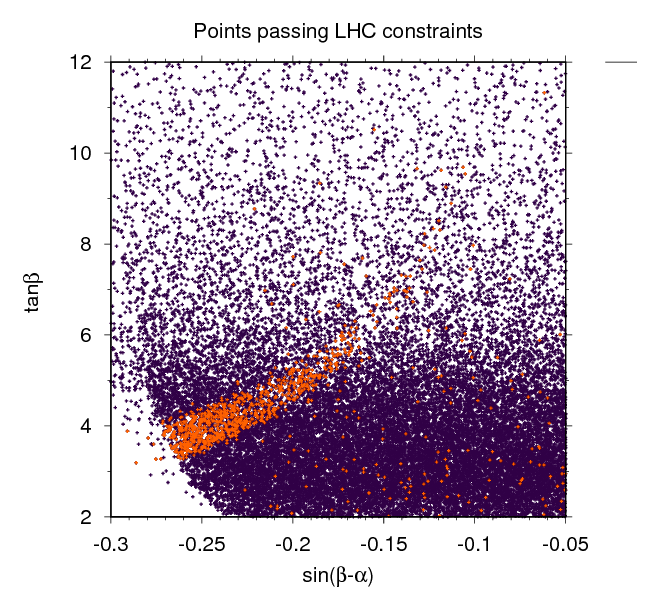}
    \end{subfigure}%
    \hfill
    \begin{subfigure}[b]{.48\linewidth}
      \includegraphics[width=\textwidth]{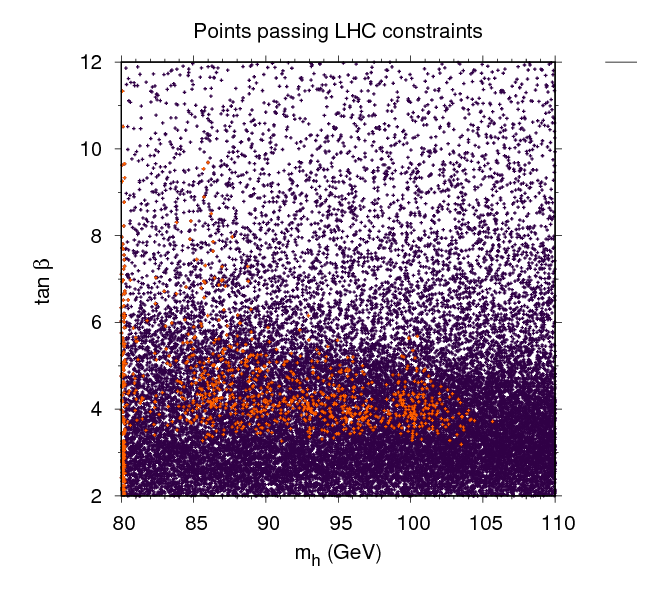}
    \end{subfigure}
\caption{Projection of the points resulting from the previous scan (see Table \ref{table:input_param2}) and passing indirect, LEP and LHC constraints in the plane $\tan\beta$ vs $\sin(\beta-\alpha)$ (left) and $\tan\beta$ vs $m_h$ (right). 
The points with a value of $\sigma_{VBF/VH}\times BR_{h\rightarrow \gamma\gamma}$ above the CMS observed 95\% C.L. upper limit are in orange; the others are in violet.}
    \label{fig:exclusion_area}
  \end{figure}

As the points above the observed CMS upper limit are excluded at 95\% C.L., we can expect to exclude some new region in the parameter space thanks to this 
analysis. To illustrate this point, in Figure~\ref{fig:exclusion_area} we plot the points resulting from the previous scan (see Table \ref{table:input_param2}) and passing the three sets of constraints in the plane $\tan\beta$ vs $\sin(\beta - \alpha)$ (left panel) and in the plane $\tan \beta$ vs $m_h$ (right panel). The violet points have a value of 
$\sigma_{VBF/VH} \times BR_{h\rightarrow \gamma \gamma}$ below the CMS observed upper limit for the corresponding mass; the orange points have a value 
of $\sigma_{VBF/VH}\times BR_{h\rightarrow \gamma \gamma}$ above the CMS observed upper limit and are consequently excluded by the experiment.

The left panel shows that most of the orange points cluster in an exclusion band in the region $\tan\beta \in$[3;6], $\sin(\beta - \alpha)\in$[-0.27;-0.14]. However, we cannot conclude that the whole orange band is excluded as we have many free parameters: the plot shows in fact a projection of a five-dimensional space  on a plane. Therefore, we can have multiple 
points with a same value of $\tan\beta$ and $\sin(\beta-\alpha)$ but with different values for the other free parameters, producing violet and orange points at the same 
position in this specific plane. Hence the orange band in the left plot of Figure~\ref{fig:exclusion_area} cannot be taken as an absolute exclusion area. 
  
\begin{figure}[h!]
    \centering
    \begin{subfigure}[b]{.42\textwidth}
      \includegraphics[width=\textwidth]{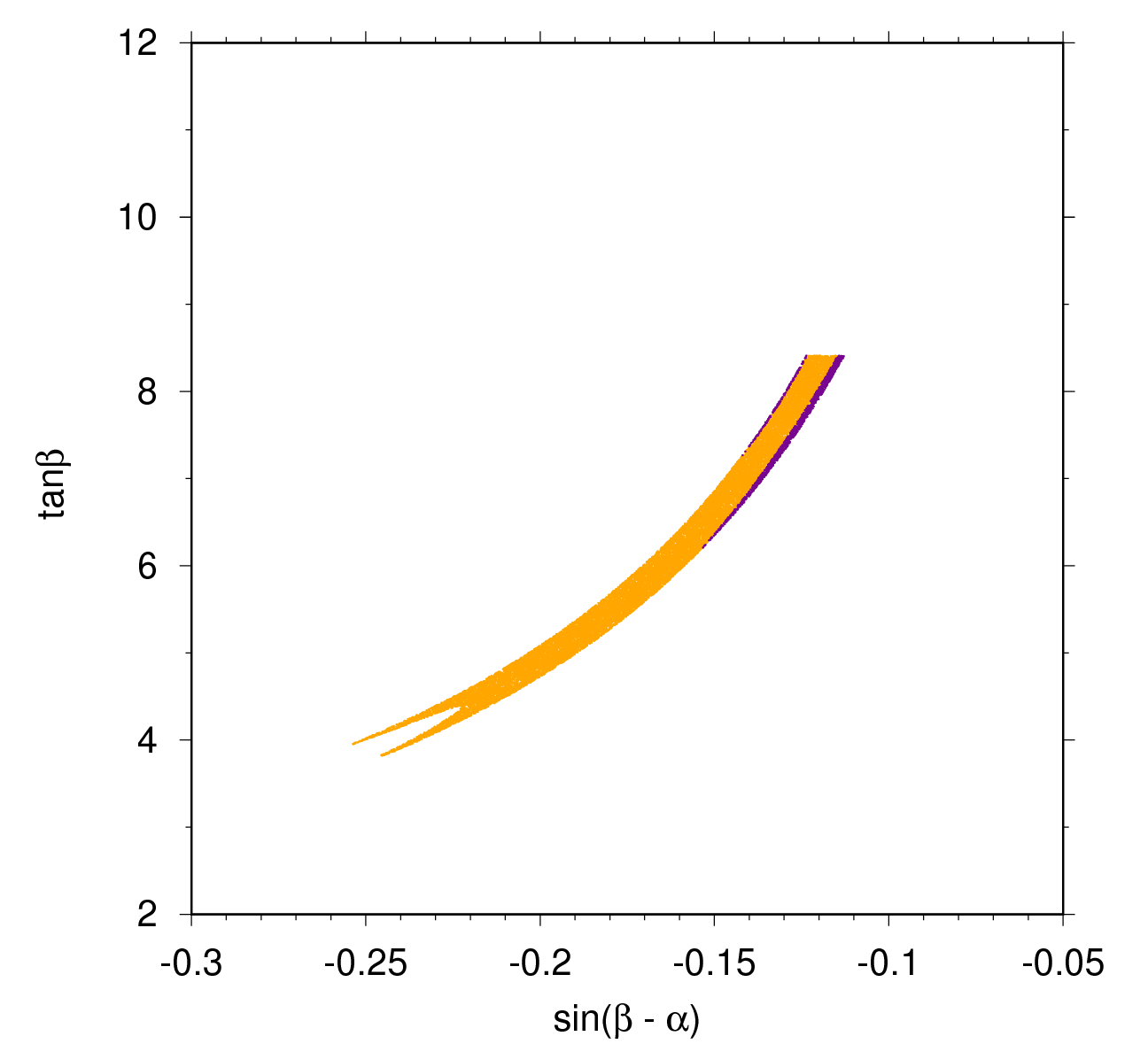}
    \end{subfigure}
    \hfill
    \begin{subfigure}[b]{.42\textwidth}
      \includegraphics[width=\textwidth]{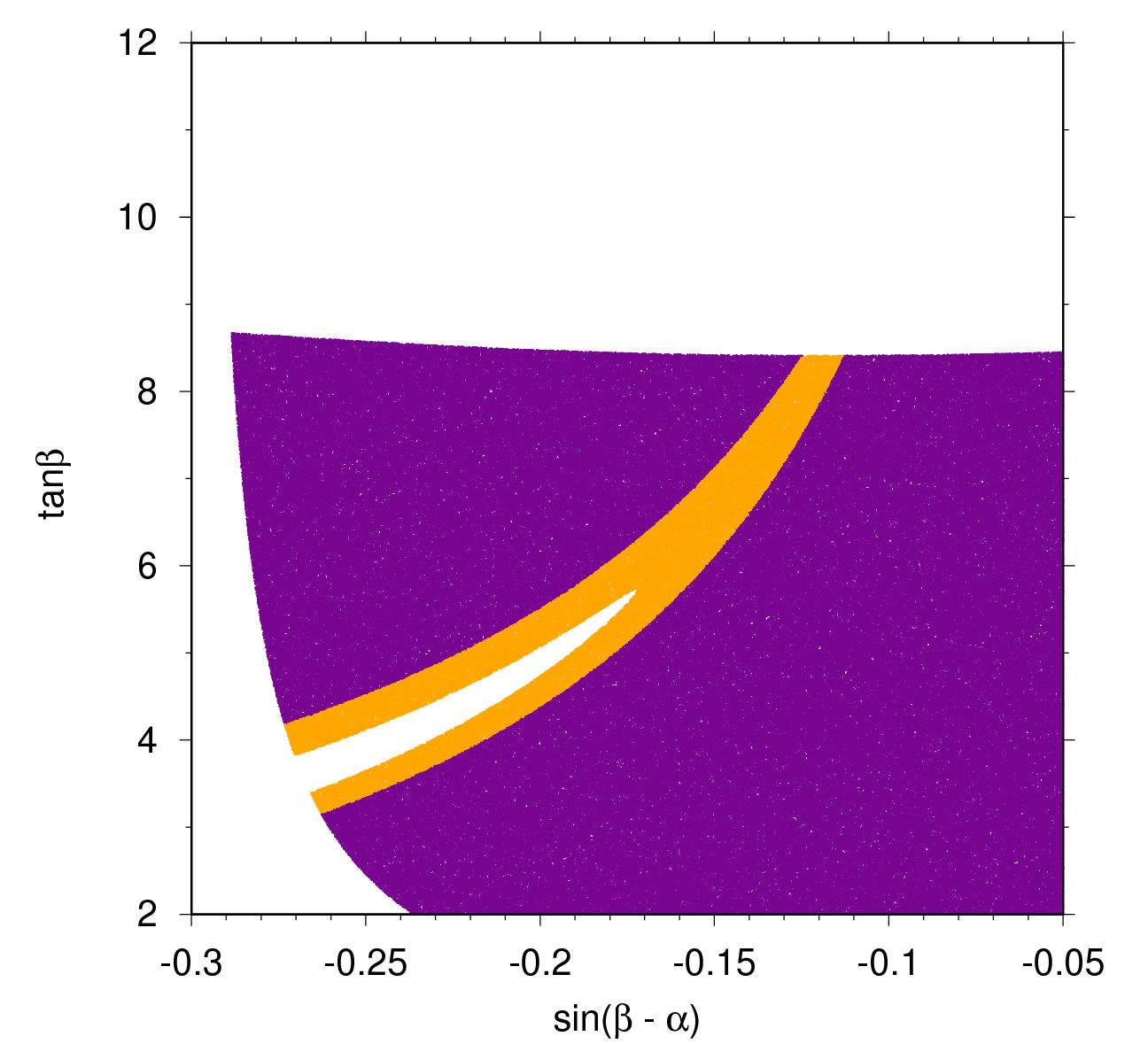}
    \end{subfigure}    
    \caption{Projection of the points passing indirect, LEP and LHC constraints in the plane $\tan\beta$ vs $\sin(\beta-\alpha)$ with $m_h=87$ GeV, $m_H$=125 GeV and $m_{12}$=30 GeV. The mass of the pseudo-scalar and charged Higgs bosons are taken to $m_A=m_{H^{\pm}}$=80 GeV (left panel) and $m_A=m_{H^{\pm}}$=500 GeV (right panel). Same color code as Figure~\ref{fig:exclusion_area}.}
      \label{fig:exclusion_area_precise_masses}
  \end{figure}

In order to illustrate this point, we produce two additional plots, shown in Figure~\ref{fig:exclusion_area_precise_masses}, in the plane $\tan\beta$ versus $\sin(\beta-\alpha)$ with 
all the other free parameters fixed. We choose $m_h=87$~GeV, $m_H$=125 GeV,  $m_{12}$=30 GeV and perform this scan for two different values 
of the mass of the pseudo-scalar and charged Higgs bosons: $m_A=m_{H^{\pm}}$=80$~$GeV (left panel) and $m_A=m_{H^{\pm}}$=500$~$GeV (right panel). As before, we only consider points passing the indirect, LEP and LHC constraints. 
The color code is the same as in Figure~\ref{fig:exclusion_area}.
The exclusion zone does not have the same shape in the two different scans and we can see that the violet points in the left panel are in orange in the panel on the right. It means that we are able to exclude some region in the plane $\tan\beta$ 
vs $\sin(\beta-\alpha)$ but the shape and extent of the exclusion zone depends on the value of the other free parameters.
  
\begin{figure}[h!]
   \centering\includegraphics[width=0.42\textwidth]{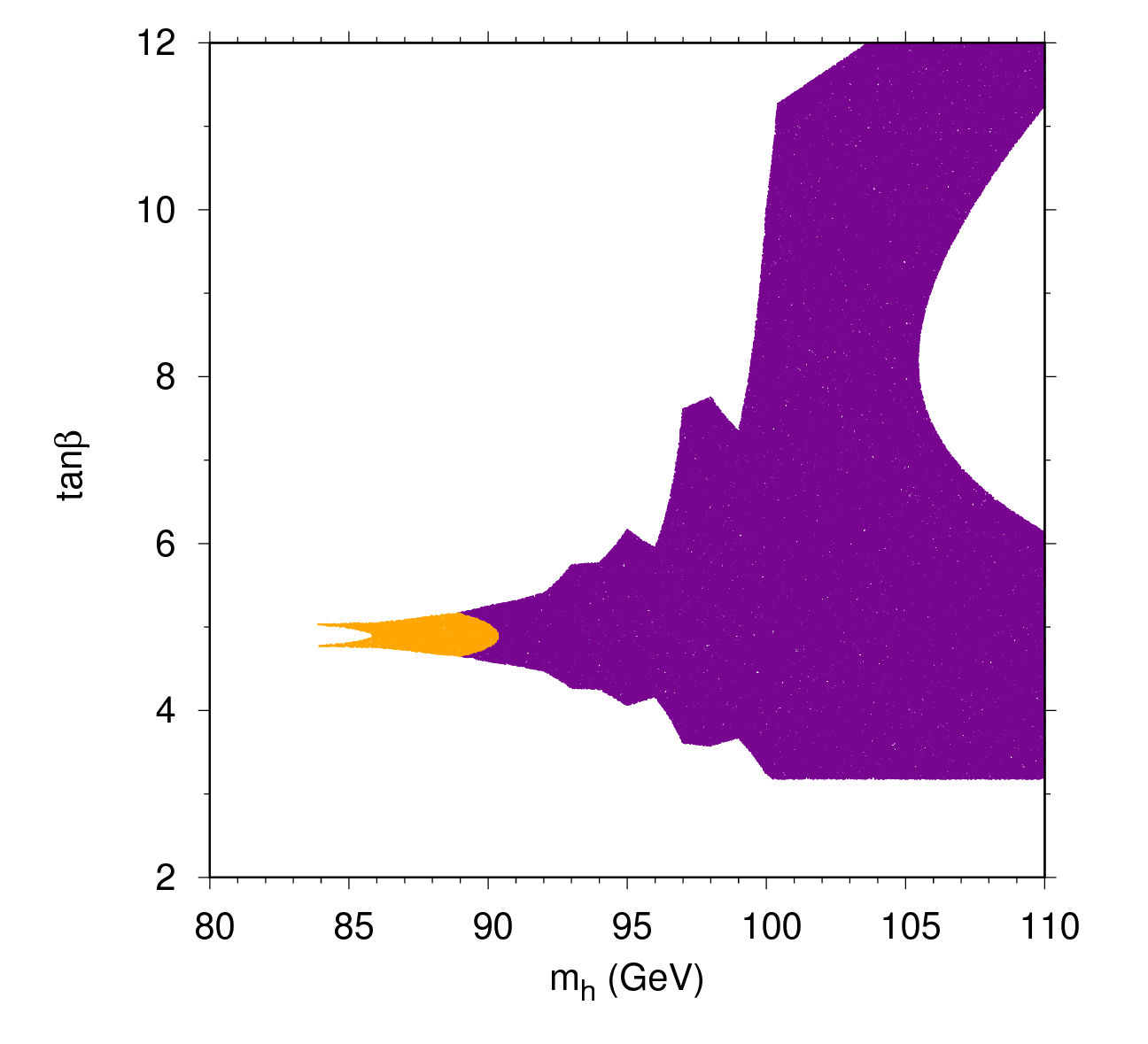}
   \caption{Projection of the points passing indirect, LEP and LHC constraints in the plane $\tan\beta$ vs $m_h$ with $m_H$=125 GeV, $m_A=m_{H^{\pm}}$=80 GeV, $\sin(\beta-\alpha)$=-0.2 and $m_{12}$=30 GeV. Same color code as Figure~\ref{fig:exclusion_area}.}
   \label{fig:exclusion_area_precise_masses_tanBmh}
  \end{figure}

 Finally, in Figure~\ref{fig:exclusion_area_precise_masses_tanBmh}, we show an exclusion zone in the plane $\tan\beta$ vs $m_h$ in the particular case where $m_H=125$ GeV, $m_A=m_{H^{\pm}}=$80 GeV, $\sin(\beta-\alpha)$=-0.2 and $m_{12}=$30 GeV. The orange points are excluded by the CMS low mass di-photon analysis at 95\% C.L..\\
  
\newpage
 \section{Search for a light pseudo-scalar Higgs boson in the 2HDMs}
 \label{sec:pseudo}
 
 In the previous section we have seen that values of the pseudo-scalar $A$ masses below 110 GeV are allowed in Type I and Lepton Specific models. It is thus natural to ask if the di-photon resonant signal may be due to the decays of the pseudo-scalar instead of the light scalar $h$.
In this section we will pursue this possibility, limiting ourselves to the same configuration studied above, i.e. fixing the mass of the heavy Higgs boson $H$ 
to $m_H=125$ GeV. The constraints on the free parameters of the model coming from indirect, LEP and LHC constraints obtained in section \ref{subsec:const_free_param} are also valid in the case of a pseudo-scalar. We can 
therefore focus on the predicted cross sections for the pseudo-scalar.
 
 As the kinematic behaviour of the two photons coming from the decay of a  pseudo-scalar particle is very similar to the the one coming 
from a scalar particle \cite{Artoisenet:2013puc}, we can directly apply the CMS study as for the scalar case to constrain a possible light pseudo-scalar. The pseudo-scalar $A$ does not 
couple at tree level to the W and Z bosons, therefore we will only focus on the gluon fusion production mode. Note also that the mass of the other light scalar $h$ is left free, and in 
principle it can also contribute to the signal at the same time as the pseudo-scalar. To simplify the analysis, however, we will not consider the possible 
bounds coming from $h$ in this case (as the available parameter space we discuss in the following gives very small cross section times branching for the pseudo-scalar $A$ which can not be probed at present).
 
 \subsection{Computation of the cross-section value}
 
The production cross section of the pseudo-scalar is different from the one for the scalar case. It is clear that for example the effective vertex with 
the gluons will be different due to different couplings and to the absence of couplings with the gauge bosons. However the ``kappa trick'' technique
used for a scalar can be used here too:
 \begin{equation}
  \sigma_{ggA}^{2HDM} \simeq \kappa_g^2\times \sigma_{ggA}^{SM}, \qquad \kappa_g^2=\frac{\Gamma_{A\rightarrow gg}^{2HDM}}
  {\Gamma_{A\rightarrow gg}^{SM}}
  \label{eq:XS_PS}
 \end{equation}
where the label $SM$ indicates that the couplings of the pseudo-scalar are set to be equal to the SM couplings of the Higgs boson (except for the different 
$\mathcal{CP}$ properties). However the values of $\sigma_{ggA}^{SM}$ are not available from the LHC Higgs Cross-Section Working Group and we cannot assume that they are the same as those of the cross section of the SM scalar 
Higgs boson $\sigma_{ggh}^{SM}$. Furthermore the program \texttt{2HDMC} does not supply the value of $\Gamma_{A\rightarrow gg}^{SM}$.
\begin{figure}[h!]
 \centering
  \includegraphics[width=0.6\textwidth]{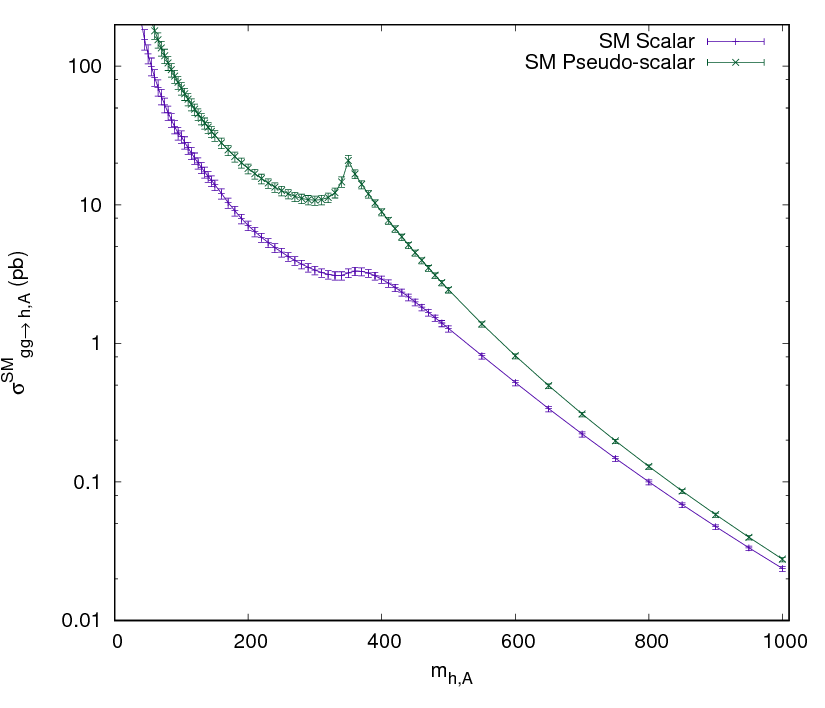}
  \caption{Production cross section in gluon fusion mode computed at NNLO by \texttt{SusHi} for an SM scalar particle (in violet) and for an SM pseudo-scalar particle (in green).}
  \label{fig:comparison_scalar_PS}
\end{figure}

We resolve the first issue by obtaining the values of the production cross-section in the gluon fusion mode for a pseudo-scalar with SM-like couplings from \texttt{SusHi} for a discrete set of values and then interpolating between the obtained values to obtain a smooth function.
Figure$~$\ref{fig:comparison_scalar_PS} shows the significant difference between the cross section obtained from \texttt{SusHi} in the gluon fusion production mode at NNLO for an SM scalar particle (in violet) and for an SM-like pseudo-scalar particle (in green) plotted as a function of the mass of the spin-0 particle.

The second issue can be overcome by using an analytical computation. The pseudo-scalar $A$ couples to the quarks as 
$g_{Aqq}=g_q^A\times i \frac{m_q}{v} \times i \gamma_5$ with $g_q^A=1$ in the SM-like case and $g_q^A=\tan\beta$ or $\cot\beta$ in the 2HDM case 
(see Table \ref{table:couplings}). The decay width of a pseudo-scalar $A$ into two gluons can therefore be written at LO as \cite{Spira:1995rr}:
\begin{equation}
 \Gamma_{A\rightarrow gg}=\frac{G_F \alpha_s^2 m_A^3}{16 \sqrt{2} \pi^3} \left| \sum_q g_q^A A_f^A(\tau_q) \right|^2
\end{equation}
with $\tau_q\equiv m_A^2/4m_q^2$ and $A_f^A$ the fermionic amplitude defined as:
\begin{equation}
 A_f^A(\tau)=\frac{f(\tau)}{\tau}, \qquad f(\tau)=\left\{\begin{array}{ll}
                                                          \arcsin^2\sqrt{\tau} & \tau \leq 1 \\
                                                          -\frac{1}{4}\left[\log\left(\frac{1 + \sqrt{1-1/\tau}}{1-\sqrt{1-1/\tau}} \right) -i\pi \right]^2 & \tau>1
                                                         \end{array}
                                                   \right. \,.
\end{equation}
The NLO corrections in the heavy top limit can be written as an additional factor to the LO width \cite{Spira:1995rr}. Considering only the top and the bottom 
quarks in the loop, we can then compute the parameter $\kappa_g^2$ at NLO in Type I:
\begin{equation}
  \kappa_g^2=\frac{\Gamma_{A\rightarrow gg}^{2HDM}}{\Gamma_{A\rightarrow gg}^{SM}}=\frac{ \left|\cot\beta \times A_f^A(\tau_t) + \cot\beta \times A_f^A(\tau_b) \right|^2}{\left|A_f^A(\tau_t) + A_f^A(\tau_b) \right|^2}\,.
\end{equation}
Using the values of $\sigma_{ggA}^{SM}$ from \texttt{SusHi} and the analytic value of $\kappa_g^2$ given above we are now able to compute the value of $\sigma_{ggA}^{2HDM}$ 
for any possible value of the free parameters.
 
\begin{figure}[h!]
 \centering
  \begin{subfigure}{0.48\textwidth}
  \includegraphics[width=\textwidth]{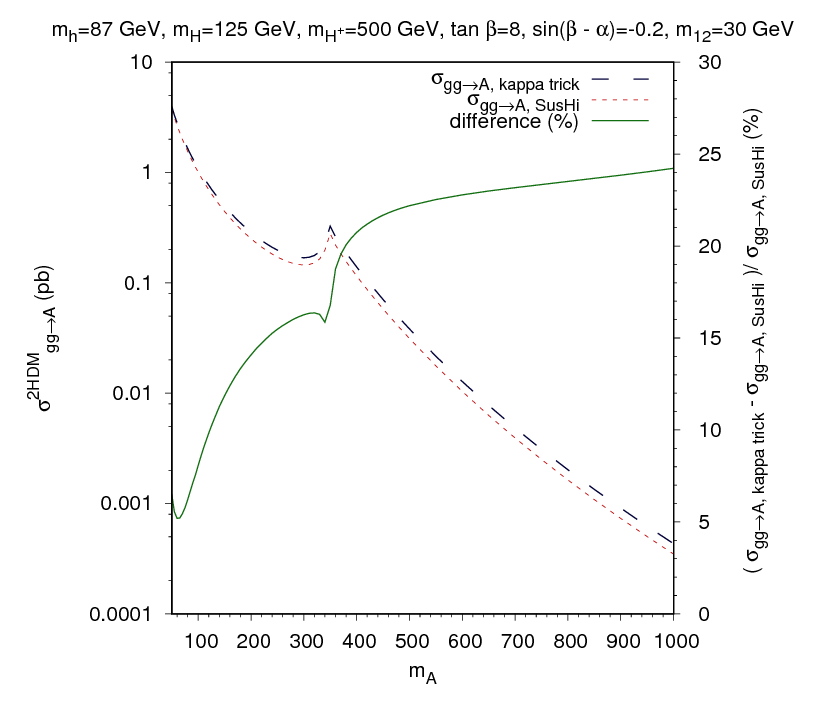}
  \end{subfigure}
  \hfill
  \begin{subfigure}{0.48\textwidth}
  \includegraphics[width=\textwidth]{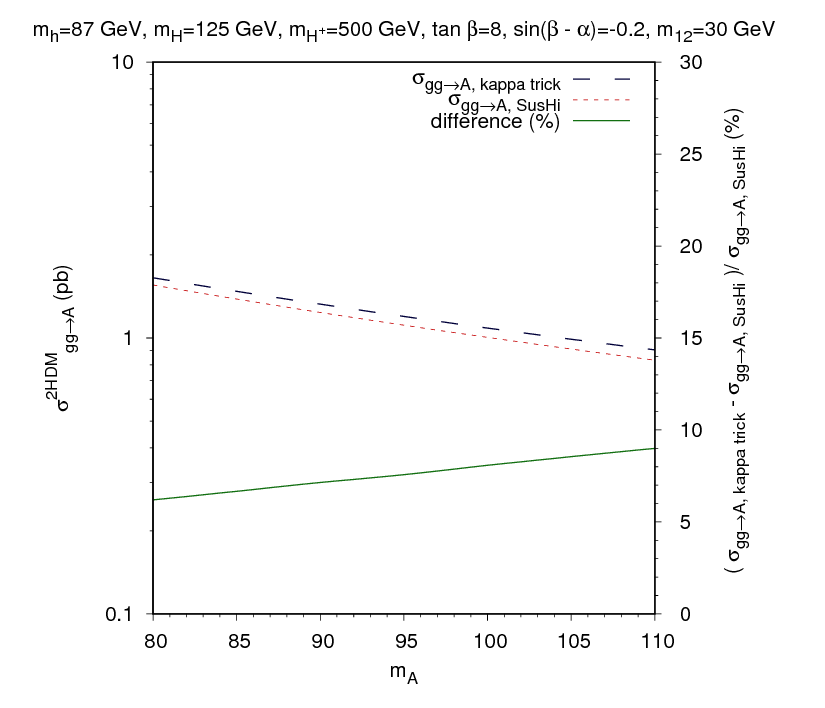}
  \end{subfigure}
  \hfill
 \caption{$\sigma^{2HDM}_{gg\rightarrow A}$ computed with the ``kappa trick'' (dashed blue line) and with \texttt{SusHi} (dotted red line). The right panel is a zoom of the left one in the low mass range.}
 \label{fig:comparison_Sushi_kappa_PS}
\end{figure}

In order to check the validity of the method,  we compare the cross-section values obtained with the ``kappa trick'' method with the 
ones given by \texttt{SusHi}. We give the results for $m_h=87$~GeV, $m_H=125$ GeV, $m_{H^{\pm}}=500$ GeV, $\tan\beta=8$, $\sin(\beta-\alpha)=-0.2$ and 
$m_{12}=30~$GeV in Figure~\ref{fig:comparison_Sushi_kappa_PS}. In the left panel the mass $m_A$ ranges from 60 GeV to 1000 GeV, while the right panel is a zoom in the 
mass region of interest for the study of a light pseudo-scalar. The dashed blue line corresponds to the cross section computed with the ``kappa trick'', the dotted red 
line to the one computed with \texttt{SusHi} and the green solid line to the deviation between the two methods.
At low mass the deviation is below 10\%, which is low enough with respect to the current uncertainties to be used in an analysis. Above $m_A=120$~GeV the 
deviation grows significantly and is about 24\% at $m_A=1000$~GeV. This is due to the NLO corrections in the ``kappa trick'' which only consider corrections in the 
infinite top mass approximation. As $m_A$ grows, this approximation becomes invalid and the cross-section value diverges from \texttt{SusHi}'s results.

 \subsection{Comparison with the CMS low mass di-photon analysis.}
 
The constraints on the free parameters coming from indirect, LEP and LHC constraints obtained in section \ref{subsec:const_free_param} remain valid for the study of a light pseudo-scalar in the scenario where the heavy scalar is identified with the SM-like one at 125 GeV. We can therefore 
perform a new scan using these bounds, with the additional constraint that the pseudo-scalar must have a mass between 80 GeV and 110 GeV in order to fit with the 
available range of the CMS analysis. The range of variation for the free parameters are given in Table \ref{table:param_scan_PS}. We restrict ourselves to Type I 
only in the gluon fusion production mode.

\begin{figure}[h!]
 \centering
  \includegraphics[width=0.6\textwidth]{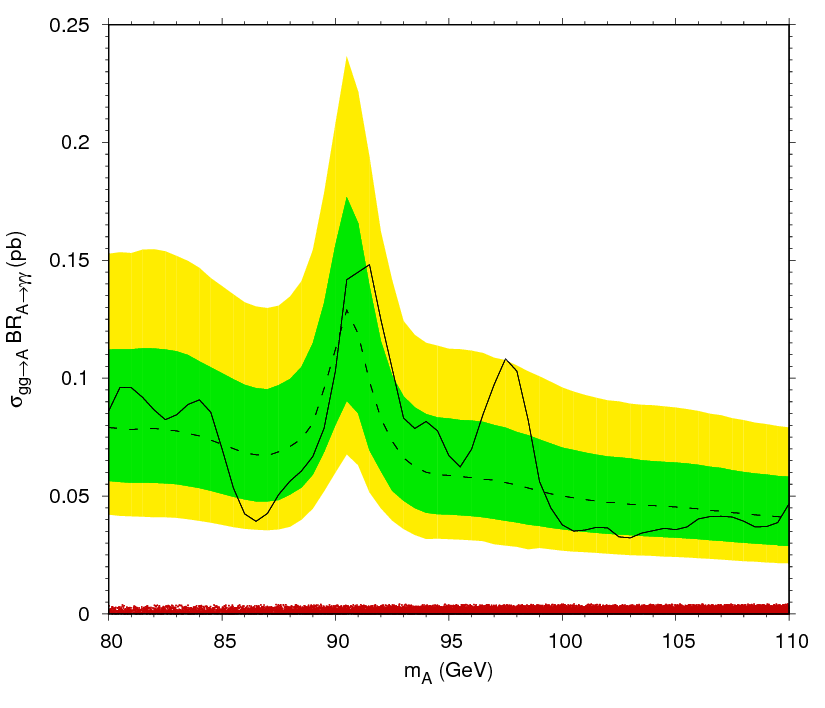}
  \caption{Points generated in the 2HDM Type I passing indirect, LEP and LHC constraints,
superimposed on the CMS 8 TeV low-mass di-photon analysis \cite{CMS_diphoton}  in the gluon fusion production mode. The dashed line corresponds to the expected upper limit at 95\% C.L..
The solid line is the observed upper limit at 95\% C.L..}
  \label{fig:XS_BR_PS}
\end{figure}

As for the scalar study we apply the indirect, LEP and LHC constraints. The resulting points are plotted in red in the plane 
$\sigma_{gg\rightarrow A}\times BR_{A\rightarrow \gamma\gamma}$ vs $m_A$ and superimposed on the CMS results in Figure~\ref{fig:XS_BR_PS}.

\begin{table}[h!]
  \begin{center}
    \begin{tabular}{c|c|c|c|c|c|c}
      $m_h$ (GeV) & $m_H$ (GeV) & $m_A$ (GeV) & $m_{H^{\pm}}$ (GeV) & $\sin(\beta-\alpha)$ & $\tan\beta$ & $m_{12}$ (GeV) \\ \hline
      [80; 110] & 125 & [80; 110] & [80; 630] & [-0.4; 0.3] & [1.5; 50] & [-(300)$^2$; +(100)$^2$]
    \end{tabular}
  \end{center}
  \caption{Range of variation for the free parameters used in the study of the pseudo-scalar A.}
  \label{table:param_scan_PS}
\end{table}

We can see that the points are well below the CMS observed upper limit on production cross section times branching ratio at 95\% C.L.. We therefore conclude that CMS had no sensitivity to a light pseudo-scalar during the LHC Run 1 in the di-photon final state.
 
 \newpage
\section{Conclusions}
\label{sec:conclusions}
The search for an extended Higgs sector is ongoing at the LHC and represents one of the most important avenues for probing the possible structure of physics beyond the Standard Model. In the simplified setting of Two Higgs Doublet Models, we have explored current constraints from flavour, precision electroweak tests and direct collider searches. We have tested
the possible reach of the CMS experiment at the  LHC Run 1 for a second Higgs particle lighter than the 125 GeV Higgs boson already discovered. We have explored in detail the different production modes (gluon fusion, vector boson fusion, 
associated production with a gauge boson) and the subsequent decay to two photons for the light boson. We have found that some sensitivity in 
these last two production modes is expected even simply recasting an existing Run 1 CMS analysis. A lighter (than the 125 GeV Higgs boson) neutral scalar or 
pseudo-scalar particle is not completely excluded by present bounds and searches. 
Out of the four types of 2HDMs, in the low-mass region for a neutral scalar, only Type I has in its parameter space 
points with large enough cross section times branching ratio to allow detection or exclusion in the gamma gamma decay channel by this analysis. We have applied this analysis also to the case of a light neutral pseudo-scalar, for which 
however cross section times branching ratio in the $\gamma\gamma$ channel is below reach at present. It is however interesting to perform such a low mass
analysis (even possibly for lower masses than those considered at Run 1) at 13 TeV for the LHC in Run 2 as the increased sensitivity to lower cross section
values will allow to further explore and constrain or possibly discover new scalar or pseudo-scalar neutral particles and in any case allow a better understanding 
of an extended Higgs sector.

 \section*{Acknowledgment}
We wish to thank Alexandre Arbey for discussions on the flavour bounds used in the present paper. 
J.T. acknowledges support from the National Natural Science Foundation of China (number 11505208) and the China Ministry of Science and Technology 
(number 2013CB838700). We also acknowledge partial support from
the Labex-LIO (Lyon Institute of Origins) under grant ANR-10-LABX-66, FRAMA (FR3127, F\'ed\'eration de Recherche ``Andr\'e
Marie Amp\`ere") and the Theory LHC-France project.

\newpage 
 \section*{Appendix}
We list in this appendix some extra numerical results in the form of plots used for the validation of the analysis.
\begin{figure}[h!]
\centering
  \begin{minipage}{0.48\textwidth}
  \centering
    \includegraphics[width=\textwidth]{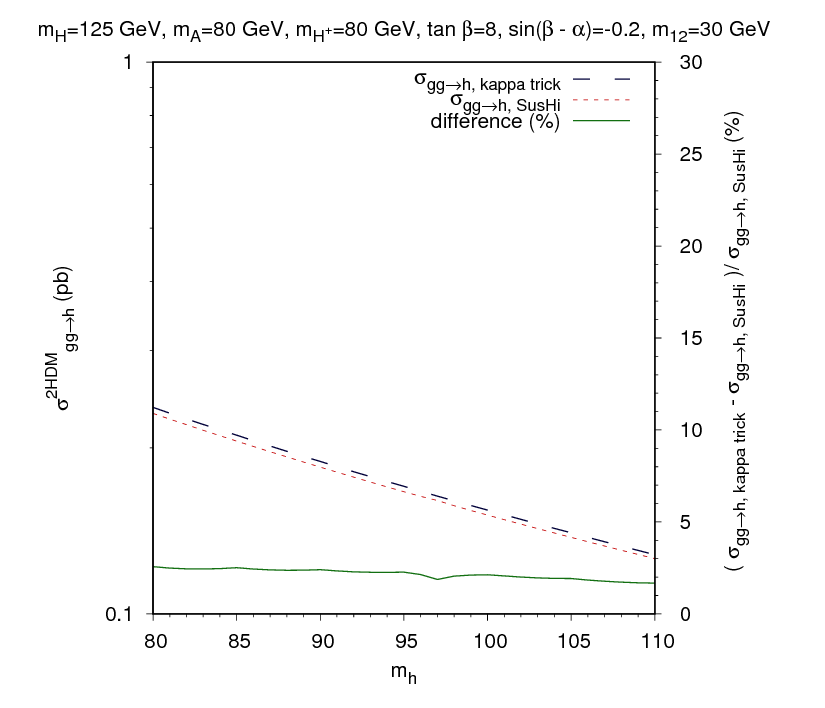}
      \caption{$\sigma^{2HDM}_{gg\rightarrow h}$  for the light Higgs boson as a function of $m_h$ computed with the ``kappa trick'' (dashed blue line) and with \texttt{SusHi} (dotted red line) and the deviation between the two (solid green line).}
      \label{fig:comparison_sushi_mA_80}
 \end{minipage}
 \hfill
 \begin{minipage}{0.48\textwidth}
    \centering
      \includegraphics[width=\textwidth]{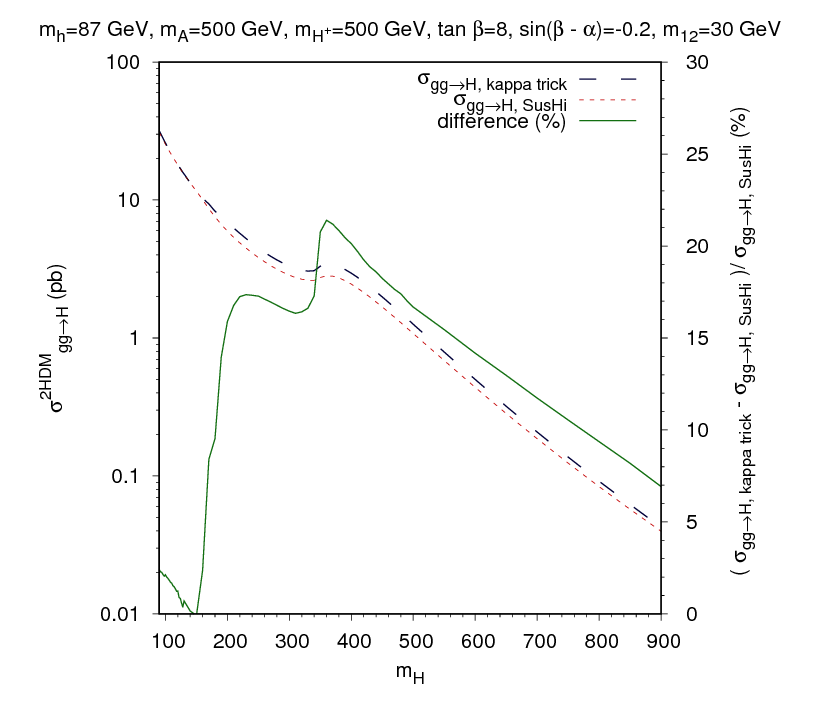}
      \caption{$\sigma^{2HDM}_{gg\rightarrow H}$  for the heavy Higgs boson as a function of $m_H$ computed with the ``kappa trick'' (dashed blue line) and with \texttt{SusHi} (dotted red line) and the deviation between the two (solid green line). As the approximation of the infinite mass for the top quark becomes false, the results with ``kappa trick'' move away from \texttt{SusHi}'s results.}
      \label{fig:XS_vs_mH}
  \end{minipage}
\end{figure}

   \begin{figure}[h!]
    \centering
      \includegraphics[width=0.6\textwidth]{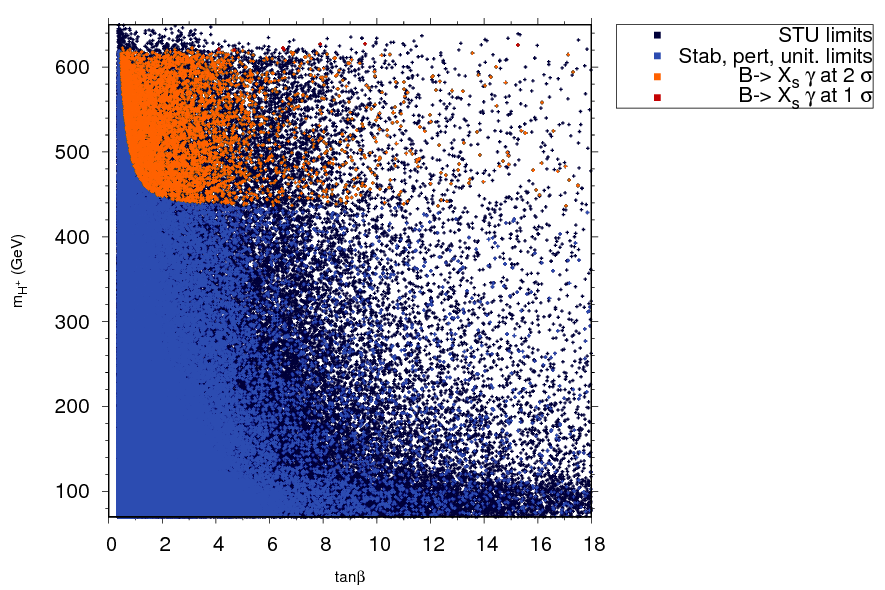}
      \caption{Points passing the S, T, U limits (dark blue), stability, perturbativity and unitarity limits (light blue) and $\mathcal{BR}(B\rightarrow X_s \gamma)$ flavor constraints at $1\sigma$ (red) and $2\sigma$ (yellow). The constraint on $\mathcal{BR}(B\rightarrow X_s \gamma)$ imposes a very hard bound on the mass of the charged Higgs bosons.}
      \label{fig:BXS_lim_Type2}
  \end{figure}

 \clearpage 
   
  \bibliography{biblio.bib}
  
\end{document}